%% file: main.tex
\documentclass[conference,nofonttune]{IEEEtran}

\usepackage{color, colortbl}
\definecolor{Gray}{gray}{0.9}

\usepackage{comment}
\usepackage[ruled,vlined,linesnumbered]{algorithm2e}
\usepackage{multirow,array,makecell} % for payoff matrix
\usepackage{wrapfig,relsize}
\usepackage{amsfonts}
\usepackage{fixmath}
\usepackage{setspace}
\usepackage{enumitem}
\usepackage{amsmath}
\usepackage{float}
\usepackage{microtype}
\input{macros}

\usepackage{mathtools,url}
\usepackage{latexsym}
\usepackage[caption=false,font=footnotesize]{subfig}
\usepackage[font={small,sf}]{caption}
\usepackage{tikz}

\usepackage[compress,sort,nospace]{cite}

\makeatletter
\g@addto@macro{\UrlBreaks}{\UrlOrds}
\makeatother

\renewcommand{\footnotesize}{\fontsize{8pt}{10pt}\selectfont}
\newcommand{\minus}{\scalebox{0.5}[1.0]{$-$}}

\newif\ifinappendix% Default is \inappendixfalse
\let\oldappendix\appendix% Store \appendix
\renewcommand{\appendix}{% Update \appendix
  \oldappendix% Default \appendix
  \inappendixtrue% Set switch to true
}

\usepackage{booktabs}

\newcommand{\rainer}[1]{{\color{red}#1}}

\newcommand{\OmitText}[1]{ {} }

\usepackage{xspace}

\newcommand{\para }[1]{\smallskip \noindent {\bf #1}}
\newcommand{\1}{{\em (i)}}
\newcommand{\2}{{\em (ii)}}
\newcommand{\3}{{\em (iii)}}

\usepackage[capitalise,nameinlink]{cleveref}

\crefname{section}{Sect.}{Sect.}
\Crefname{section}{Section}{Sections}

\usepackage{url}
\makeatletter
\g@addto@macro{\UrlBreaks}{\UrlOrds}
\makeatother
\author{
    \IEEEauthorblockN{George Bissias\IEEEauthorrefmark{1}, Rainer Böhme\IEEEauthorrefmark{2}, David Thibodeau\IEEEauthorrefmark{3}, Brian N. Levine\IEEEauthorrefmark{1}}
    \IEEEauthorblockA{\IEEEauthorrefmark{1}University of Massachusetts Amherst
    \\\{gbiss, brian\}@cs.umass.edu}
    \IEEEauthorblockA{\IEEEauthorrefmark{2}University of Innsbruck
    \\rainer.boehme@uibk.ac.at}
    \IEEEauthorblockA{\IEEEauthorrefmark{3}Unaffiliated
    \\davidpthibodeau@gmail.com}
}

\begin{document}

\title{Pricing Security in Proof-of-Work Systems}

\maketitle

\begin{abstract}
\em A key component of security in decentralized blockchains is proof of opportunity cost among block producers. In the case of proof-of-work (PoW), currently used by the most prominent systems, the cost is due to spent computation. In this paper, we characterize the security investment of miners in terms of its cost in fiat money. This enables comparison of security allocations across PoW blockchains that generally use different PoW algorithms and reward miners in different cryptocurrency units. We prove that there exists a unique allocation equilibrium, depending on market prices only, that is achieved by both strategic miners (who contemplate the actions of others) and by miners seeking only short-term profit. In fact, the latter will unknowingly compensate for any attempt to deliberately shift security allocation away from equilibrium.

Our conclusions are supported analytically through the development of a Markov decision process, game theoretical analysis, and derivation of no arbitrage conditions. We corroborate those results with empirical evidence from more than two years of blockchain and price data. Overall agreement is strong. We show that between January 1, 2018 and August 1, 2020, market prices predicted security allocation between Bitcoin and Bitcoin Cash with error less than 0.6\%. And from the beginning of October 2019, until August 1, 2020, market prices predicted security allocation between Bitcoin and Litecoin with error of 0.45\%. These results are further corroborated by our establishment of Granger-causality between change in market prices and change in security allocation.

To demonstrate the practicality of our results, we describe a trustless oracle that leverages the   equilibrium to estimate the price ratios of PoW cryptocurrencies from on-chain information only. 
\end{abstract}

\setlength{\belowdisplayskip}{3pt} 
\setlength{\belowdisplayshortskip}{3pt}
\setlength{\abovedisplayskip}{3pt} 
\setlength{\abovedisplayshortskip}{3pt}

\input{introduction}
\input{background}
\input{opt_alloc}

\input{framework}
\input{nash}

\input{na}

\input{evaluation}

\input{econometrics} 
\input{oracle}

\input{related}
\input{conclusion}

\clearpage
\pagebreak

\urlstyle{sf}
\pagestyle{plain}
{\footnotesize \bibliographystyle{IEEEtran}
\bibliography{IEEEabrv,references}}

\appendices

\input{symbols}
\input{suppl_figs}
\input{mdp_details}
\input{example}
\input{algorithms}
\input{proofs}

\end{document}

%% file: macros.tex
\usepackage[framemethod=tikz]{mdframed}
  
\surroundwithmdframed[
  skipabove=0em,
  skipbelow=0ex,
  outerlinewidth=0.4pt,
  innerlinewidth=0.4pt,
  middlelinewidth=1pt,
  middlelinecolor=white,
  bottomline=false,topline=false,rightline=false]{mythm}

\surroundwithmdframed[
  skipabove=1em,
  skipbelow=0ex,
  outerlinewidth=0.4pt,
  innerlinewidth=0.4pt,
  middlelinewidth=1pt,
  middlelinecolor=white,
  bottomline=false,topline=false,rightline=false]{mylem}

\surroundwithmdframed[
  skipabove=0em,
  skipbelow=0ex,
  outerlinewidth=0.4pt,
  innerlinewidth=0.4pt,
  middlelinewidth=1pt,
  middlelinecolor=white,
  bottomline=false,topline=false,rightline=false]{mycor}

\newcounter{myprot}
\newcounter{myalg}
\newcounter{myprop}
\newenvironment{myprop}
{\refstepcounter{myprop}  \noindent \textbf{PROPOSITION \arabic{myprop}:}}
{\vspace{.25em}}

\newcounter{myprob}
\newcounter{myexam}
\newenvironment{myexam}
{\refstepcounter{myexam}  \noindent \textbf{EXAMPLE \arabic{myexam}:}}
{\vspace{.25em}}

\newcounter{mythm}
\newenvironment{mythm}
{\refstepcounter{mythm}  \noindent \textbf{THEOREM \arabic{mythm}:}\em}
{\vspace{.25em}}

\newcounter{mylem}
\newenvironment{mylem}
{\refstepcounter{mylem}  \noindent \textbf{LEMMA \arabic{mylem}:}\em}
{\vspace{.25em}}

\newcounter{mycor}
\newenvironment{mycor}
{\refstepcounter{mycor} \vspace{1em} \noindent \textbf{COROLLARY \arabic{mycor}:} \em}
{\vspace{.5em}}

\newcounter{myobs}
\newcounter{mydef}
\newenvironment{mydef}
{\refstepcounter{mydef} \vspace{1em} \noindent \textbf{DEFINITION \arabic{mydef}:}}
{\vspace{.5em}}

\newcounter{myconj}
\newenvironment{myproof}
{\noindent \textbf{PROOF:}}
{\vspace{-3ex}\begin{flushright} $\Box$ \end{flushright}\vspace{2ex}}

%% file: introduction.tex
% !TEX root = main.tex

\section{Introduction}

Cryptocurrencies such as Bitcoin~\cite{Nakamoto:2009} have emerged as an intriguing complement to state-backed fiat currencies. 
We analyze the security of blockchains in the form of distributed systems with decentralized control and weak identification of participants (i.e., Sybil attacks~\cite{Douceur:2002} are possible and require mitigation). 
 
Typically, cryptocurrencies are implemented using a blockchain data structure, with each block containing a set of transactions.
Although there are many aspects of blockchain security, in this paper we focus on security as it relates to consensus on the contents of blocks. 

A blockchain is secure only to the extent that consensus emerges from the entire set of participants rather than an individual or subgroup.
For blockchains with open membership, consensus is based on one of several different mechanisms including \emph{proof-of-work (PoW)}~\cite{Nakamoto:2009} and \emph{proof-of-stake (PoS)}~\cite{King:2012}, which are the two most popular choices. To gain the authority to record transaction history to the blockchain, the former forces participants to demonstrate use of computational resources, while the latter requires that participant funds be locked for a fixed period of time. Because participants could invest their resources elsewhere, participation in consensus, and thus the basis of blockchain security, is the summed opportunity cost of all participants. This cost is offset by a reward paid in cryptocurrency, which has a market-driven fiat value. Blockchains are secure when the opportunity cost cannot be borne by one dishonest participant or group that seeks to control consensus. 
Thus, the relative security of cryptocurrencies can be determined from the fiat value of their opportunity costs. 

The relationship between the resources that participants choose to allocate among chains and the market-based fiat exchange value of each cryptocurrency is fundamental to the amount of security provided, and yet it is not well understood. In this paper, we provide novel analysis of this relationship and the allocation 
 chosen by participants to one blockchain over another.  
Earlier work offers an incomplete understanding of this relationship. Spiegelman et al.~\cite{Spiegelman:2018} predicted the existence of stable equilibria among resource allocations, and Kwon et al.~\cite{Kwon:2019} subsequently identified multiple Nash equilibria, with one being closely observed in practice. Indeed, there exists evidence that some blockchain participants are already aware of this equilibrium~\cite{Zhuoer:2020}.

Although Kwon et al.~\cite{Kwon:2019} take an important first step, we feel their analysis requires refinement. Their utility function is not parsimonious, relying on multiple miner-types unnecessarily so that the regime for each equilibrium cannot be determined without unobservable information. This precluded critical tests of the theory in their work. The following questions remain unanswered in their work. Why does one equilibrium dominate all others in practice? How do nonstrategic agents \emph{find} this equilibrium. And how does the equilibrium change with protocol details such as cryptocurrency inflation rate or choice of PoW algorithm? These questions are critical to understanding PoW blockchain security, and to the best of our knowledge, the present work is the only one that provides formal answers. 

A secondary goal of our work is to bridge the gap between techniques familiar to computer scientists and those more commonly applied in the field of economics and finance. We believe that both communities can benefit from this synthesis. On one hand, it has been argued~\cite{Bohme:2015} that cryptocurrencies provide a real-world, highly transparent, and greatly simplified testbed for financial and economic theories. 
And on the other hand, the study of economics and finance provides computer scientists deeper insight into the behavior of agents who operate in the presence of monetary incentives. In this paper, we demonstrate how the latter can be used to develop a robust and highly accurate theory of security in PoW blockchains.
For example, we show that between January 1, 2018 and August 1, 2020, cryptocurrency prices alone were sufficient to predict resource allocation between Bitcoin and Bitcoin Cash with root mean squared error 
0.59\%; during overlapping periods, this error is three times lower than prior work.
And beginning October 2019, until August 1, 2020, we show that a combination of cryptocurrency and hash price data is sufficient to predict resource allocation between Bitcoin and Litecoin (which do not share a PoW algorithm) with error of 0.45\%.

The implications of our findings are profound for the blockchain ecosystem. They provide insight into the motivations and reasoning of PoW \emph{miners}, the typically reticent participants responsible for securing blockchains. For at least the past two years, the scope of our data, change in currency price alone has proven to be a remarkably accurate predictor of change in miner resource allocation. Our results suggest that this connection is Granger-causal~\cite{Granger:1969}: changes in the fiat value of a cryptocurrency will tend to result in a rapid change to investment in its security. 

In sum, we make the following contributions.

\begin{enumerate}[leftmargin=14pt]
\item We use a multi-method approach that spans solution concepts established in computer science, economics, and finance. Specifically, we use a Markov decision process (MDP) to analyze basic resource allocation dynamics, competitive game theory for multi-miner interaction, and consolidate our theory in no-arbitrage conditions~\cite{LeRoy:2014}, a powerful solution concept in finance that is more general than MDPs and requires fewer assumption than Nash equilibria within game theory. This analysis yields a single equilibrium allocation that we show to be an attractor; every other allocation will tend to rebalance toward it.
\item We evaluate the strength of this attractor on more than two years of historical blockchain and price data for many of the most popular PoW blockchains including Bitcoin, Ethereum, Bitcoin Cash, Ethereum Classic, and Litecoin. We show that actual resource allocation among blockchains that share the same PoW algorithm follows extremely close to the equilibrium; those that do not share a PoW algorithm also follow closely, but less so due to market inefficiencies. 
\item Using Granger-causality, we show that, on a systematic, hourly basis, change in the fiat value of a cryptocurrency elicits a change in the resources a miner allocates to securing its blockchain. We also show that the opposite link is generally rare, but has manifested during tumultuous historical events.
\item We leverage the correlation between actual and predicted resource allocations to describe how to develop a trustless exchange price ratio oracle between any pair of PoW cryptocurrencies. Its susceptibility to manipulation is limited relative to other decentralized solutions. And, to the best of our knowledge, it is the first to use only on-chain information in a way that allows for quantification of manipulation cost.
\end{enumerate}
We conclude with a comparison to related work. 

%% file: background.tex
% !TEX root = main.tex

\section{Security in PoW Blockchains}
\label{sec:background}

A distributed and decentralized \emph{blockchain}~\cite{Nakamoto:2009} (or \emph{chain} for brevity) is a data structure, formed among autonomous peers having weak identities~\cite{Douceur:2002}, who assemble \emph{blocks} in a hash-linked list. Each block contains a set of \emph{transactions}, which can be simple account updates or more complex state changes in smart contracts. Transactions are \emph{confirmed} once they appear in a block on the chain. 

\para{Mining.} PoW \emph{miners} achieve consensus through a block \emph{mining} process. The purpose of the mining process is to compensate for the absence of strong identities by requiring each peer participating in the blockchain to provide evidence of computation. In its simplest form, each miner applies a cryptographic hash algorithm~\cite{Back:2002} to the metadata associated with the block called the \emph{block header}, randomly varying a nonce in that header. 
If the resulting hash value is less than a known \emph{target}, then the miner is considered to have mined the block and it is awarded a portion of cryptocurrency (or \emph{coin} for brevity): some is newly minted to form a base reward and the rest is derived from transaction fees. Coins carry an exchange value in \emph{fiat currency} (a state-backed currency such as USD), which is established by exchanges that facilitate trade.
\emph{Difficulty} is a quantity inversely related to the target by a constant. It is essentially the expected number of hashes required to mine a block, and we treat it as such unless otherwise indicated. (Most blockchains actually define the difficulty somewhat differently, but our definition is similar in spirit.)
 The difficulty (and therefore the target) is updated via a protocol-defined algorithm called a \emph{difficulty adjustment algorithm} (DAA) so that all miners, working independently, are expected to mine a block in a fixed expected time (e.g. 600 seconds in Bitcoin). 

\para{Threat model.} Blockchain security is multifaceted~\cite{Li:2020}; vulnerabilities can arise at the network~\cite{Biryukov:2014,Heilman:2015}, protocol~\cite{Porta:2019}, consensus~\cite{Nakamoto:2009,Eyal:2014}, or application~\cite{Castillo:2016} layers. But perhaps the most fundamental attack on PoW blockchains is the \emph{51\% attack}, which arises when the computational resources of a nefarious individual or organization exceed those of the remaining participants. In this work, as is common in related works~\cite{Kwon:2019,Eyal:2014}, we assume that attackers cannot break primitives or exploit network or cryptographic vulnerabilities and that they have potentially substantial but ultimately limited resources. Because attacker hash rate is assumed to be limited, risk of a 51\% attack is lowest on blockchains where absolute hash power is highest~\cite{Rosenfeld:2012,Sapirshtein:2015}.

\begin{mydef}
The \emph{security metric} for PoW blockchains is hash rate.
\end{mydef}

\vspace{-0.5em}

\para{Hash markets.} PoW mining constitutes a bona fide cost to miners in terms of both capital outlay and expended electricity~\cite{Hass:2018}. The majority of work performed on all major PoW blockchains uses application specific integrated circuits (ASICs). Purchasing ASICs constitutes a significant capital expenditure and also creates lock-in because these devices can typically only be used to execute a single PoW algorithm. Yet some blockchains, such as Bitcoin and Bitcoin Cash, use the same PoW algorithm. In this case, the cost to move mining resources to the other chain is negligible. This creates an economic tension between such blockchains whereby the incentive to mine on a given chain vacillates depending on the relative fiat value of reward per hash at any given moment. Moreover, there exist markets~\cite{nicehash} for renting time on ASICs, which allow miners to effectively purchase reward on blockchains implementing a PoW algorithm that they cannot mine directly themselves, or sell excess capacity and thus amortize capital they have invested in ASICs.

%% file: opt_alloc.tex
% !TEX root = main.tex

\section{A Motivating Example}
\label{sec:motivation}

In this section, we illustrate by example how any miner given the choice between two blockchains will allocate his hash rate to each in proportion to its share of the total reward  to optimize his profit. This principle is carried forward throughout the paper.

Imagine a simplified blockchain ecosystem where there exist only two chains $A$ and $B$, each implementing the same PoW algorithm and each aiming to produce blocks at the same average rate of $T=2$ seconds. The coins issued by $A$ carry 2 units of fiat value while those issued by $B$ carry only 1 unit. There exists a single miner who must decide how to allocate his available hash rate of $H = 6$ hashes per second among the two chains so as to maximize profit. We assume that each chain's DAA fully adjusts to the hash rate applied to that chain after a single block. 

Suppose that initially the miner's hash rate is split evenly among the two chains. What is the miner's optimal hash rate \emph{rebalancing} given the initial difficulty on each chain?
To answer this question, we use a Markov decision process (MDP). States in the MDP correspond to difficulty associated with each chain, which we measure in terms of the expected number of hashes required to mine a block. Actions correspond to the miner's hash rate \emph{allocation} among the two chains. And transitions occur from one state to another when a block is mined using the hash rate given by the current action. Further MDP details can be found in Appendix~\ref{sec:mdp_details}. Figure~\ref{fig:mdp_a6} shows the optimal policy for the miner in a grid where each column represents the current difficulty on chain $A$ and each row represents the current difficulty on chain $B$. The direction of the arrow at each grid point indicates the direction of optimal hash rate rebalancing among the two chains. 

\begin{figure}[t]
\centerline{\includegraphics[trim=20 5 20 40, clip,width=.9\linewidth]{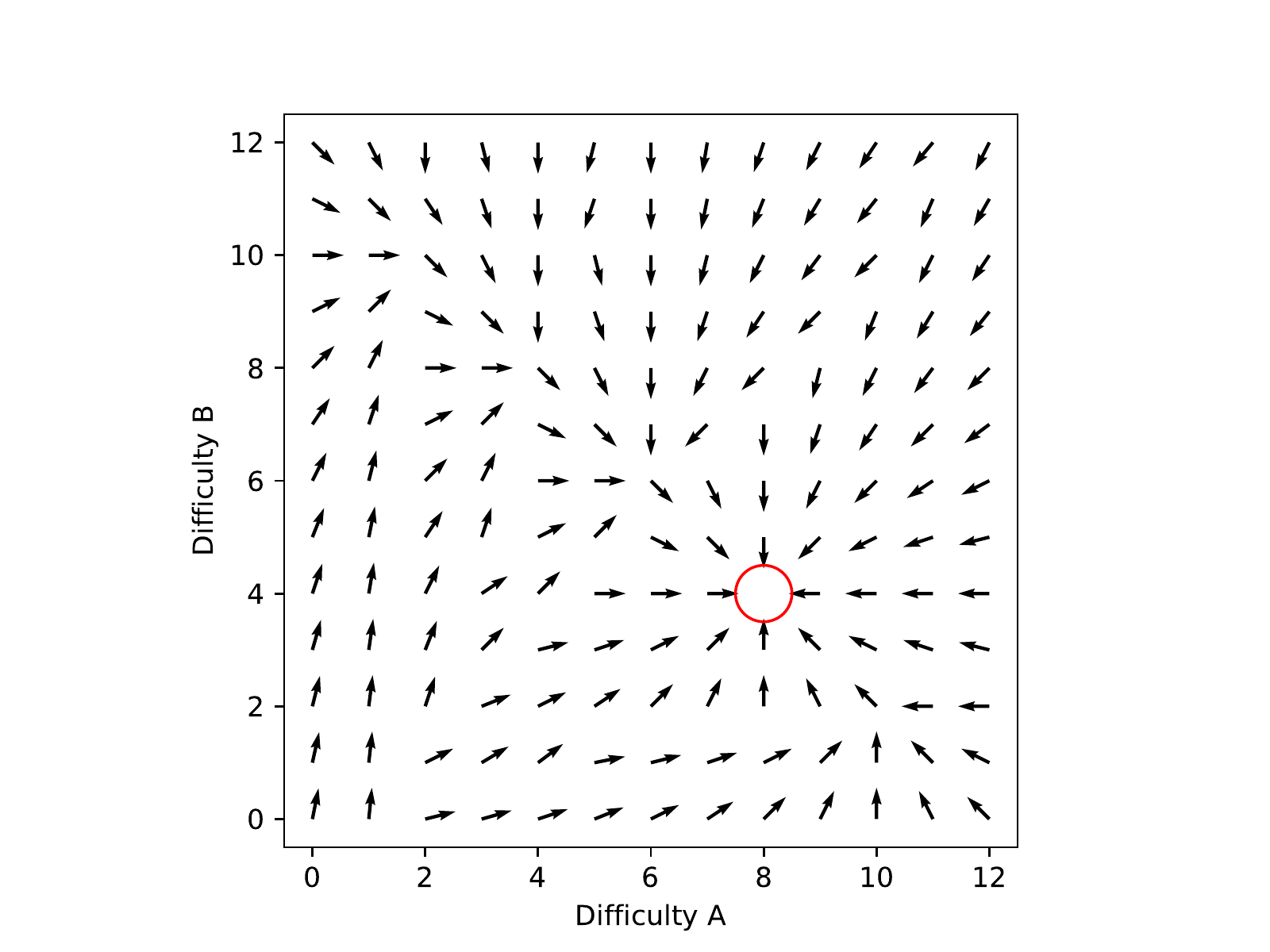}}
\caption{\emph{Direction of optimal allocation of a total of 6 hashes per second among chains $A$ and $B$ for given difficulties (expected hashes per block). Each point in the plot represents difficulties for chains $A$ and $B$. The direction of arrows indicates how to optimally rebalance allocation at the given difficulties. The concentration point for this miner, the only state where remaining stationary is optimal, is indicated by the red circle.}}
\label{fig:mdp_a6}
\end{figure}

The figure reveals that the optimal action for the miner is not simply to allocate all hash rate to chain $A$, which offers the highest reward per block. This is because shifting all hash rate to $A$ forces the difficulty higher, raising cost. \emph{This tension captures the essence of an equilibrium that forms between blockchains.}

The main diagonal of the grid (from upper left to lower right) in Figure~\ref{fig:mdp_a6} corresponds to states where the current aggregate difficulty (across both chains) is equivalent to the total hash rate that the miner can apply to the chains during the time it takes to generate a block. Notice that the difficulties on this diagonal always sum to 12 because blocks are generated every 2 seconds on average, and the miner can generate 6 hashes per second. We see two major phenomena emerge. First, when the aggregate difficulty across the two chains  requires a hash rate that differs from what is possessed by the miner (off-diagonal regions), the optimal action for the miner is to move toward the main diagonal. Second, along the main diagonal, the optimal action is always to move toward the state indicated by the red circle, which corresponds to a difficulty of eight hashes per block for chain $A$ and four hashes per block for chain $B$. This is the only state for which the optimal action is to remain stationary, making it the only concentration point in the grid. 

Thus, it appears that \emph{the optimal strategy is for the miner to allocate his hash rate on each chain in proportion to the chain's share of the total reward}. The MDP confirms that this property holds for other reward proportions, so it does not appear to be a coincidence.
This result implies that the fiat value of a blockchain's native coin has a direct and quantifiable impact on its security relative to another blockchain. In the remainder of this document we explain why this is the case and both generalize and formalize the result. 
Going forward, we model multiple competing miners who allocate resources among blockchains generally having different block times and PoW algorithms. In this broader context, we observe the formation of an allocation \emph{equilibrium} that forms at precisely the same point achieved by the lone miner in this example.

%% file: framework.tex
% !TEX root = main.tex

\section{Framework}
\label{sec:framework}

In this section, we construct an analytical framework, which generalizes familiar blockchain concepts and enables presentation of novel concepts in subsequent sections.

We consider two blockchains $A$ and $B$, each generally using different PoW algorithms $\texttt{ALG}_A$ and $\texttt{ALG}_B$. Having different PoW algorithms, we imagine that the sets of miners $M_A$ and $M_B$ of each coin are generally disjoint, but in the special case where $\texttt{ALG}_A = \texttt{ALG}_B$ or when the algorithms are supported by the same mining hardware, their intersection can be non-empty. We denote the set of all miners by $M = M_A \cup M_B$. 

We denote the \emph{hash rate} (hashes per second) for miner $m$ by $H(m)$, and with $H_A$ and $H_B$ we denote the \emph{aggregate} hash rate of all miners on chains $A$ and $B$, respectively. We assume that the hash rate for each miner remains constant over time as does the total hash rate $H$. Through secondary \emph{hash rate markets} such as NiceHash~\cite{nicehash}, it is possible for a miner $m \in M_A$ to trade hash power  in $A$ (through a series of trades) for hash power in $B$. Thus, the sets $M_A$ and $M_B$ are fluid, i.e. miners can readily move between sets. 

We denote by $T_X$ the \emph{target average block inter-arrival time} for chain $X \in \{A, B\}$. In general, blocks from chains $A$ and $B$ will be produced at different times, but we require some method of marking time universally. Let $\tau$ be a discrete variable that represents the times when a block is mined on chain $A$ or $B$. At time $\tau$, the \emph{actual} inter-arrival time for the latest block from chain $X$ is given by $t_X(\tau)$. 

\begin{mydef}
\label{def:spot_hashes}
The \emph{spot hash price} at time $\tau$, denoted $\sigma_X(\tau)$, is the fiat price of a single hash using PoW algorithm $\texttt{ALG}_X$. 
\end{mydef}

The spot hash price on a given blockchain is simply the cost to purchase hash power on that chain. Using Definition~\ref{def:spot_hashes}, and assuming a perfectly efficient hash rate market, we can quantify the fiat value of hash power devoted to securing a given blockchain.

\begin{mydef}
\label{def:chain_security}
 The \emph{actual security investment} in blockchain $X$, denoted $s_X(\tau) = H_X(\tau) \sigma_X(\tau)$, is the actual fiat value of hash power devoted by miners to mining on chain $X$ for one second at time $\tau$. 
\end{mydef}

The actual security investment definition abstracts the conventional concept of hash rate by converting hashes per second to fiat per second.

\begin{mydef}
\label{def:rel_sec}
The \emph{security allocation} at time $\tau$, denoted by vector $\boldsymbol{w}(\tau)$, is the fraction of the total actual security investment applied to each chain:
\begin{equation}
\label{eq:security}
\boldsymbol{w}(\tau) = (w_A(\tau), w_B(\tau)) = \frac{1}{s_A(\tau) + s_B(\tau)} (s_A(\tau), s_B(\tau)).
\end{equation}
\end{mydef}
\vspace{-1em}

We often refer to a security allocation as simply an allocation for brevity, and we also drop $\tau$ from the notation when time is either unimportant or clear from context. Notice that the security allocation to chain $X \in \{A, B\}$ is equivalent to the fraction of total actual security investment allocated to chain $X$. 
Thus, when chains $A$ and $B$ share the same PoW algorithm, $\sigma_A = \sigma_B$ and $\boldsymbol{w}$ gives the share of total hash rate allocated to each chain. At times we consider the relative security only for miner $m_i$, which we denote by $\boldsymbol{w}_{\!i} = (w_{iA}, w_{iB})$, where 
\begin{equation}
\sum_{i, m_i \in M} \boldsymbol{w}_{\!i} = \boldsymbol{w}.
\end{equation}

The fiat value of the coinbase reward plus average fees for chain $X$ is given by $V_X(\tau)$. Coinbase value decomposes into $V_X(\tau) = k_X(\tau) P_X(\tau)$, where $k_X(\tau)$ is the quantity of $X$ coins (from base reward and average transaction fees) paid out per block, and $P_X(\tau)$ is the fiat value of each coin from chain $X$ at time $\tau$. 

Finally, define the relative reward of the two chains by $R(\tau) = \frac{V_A(\tau)}{V_A(\tau) + V_B(\tau)}$. 

\begin{mydef}
The \emph{target security investment} $S_X(\tau)$ for a blockchain $X$ at time $\tau$ is the fiat value of hash power that must be applied to chain $X$ by miners, for each second beginning at time $\tau$, to produce a block in expected time $T_X$.
\end{mydef}

Recall from Section~\ref{sec:background} that the difficulty of a blockchain measures the expected number of hashes required to mine a block. 
The target security investment abstracts the difficulty by converting hashes per block to fiat per second.
Contrast actual and target security investments $s$ and $S$ with conventional quantities hash rate and difficulty $H$ and $D$. Quantities $s$ and $H$ are controlled by the miner, they reflect actual resources devoted to mining, while quantities $S$ and $D$ are set by the blockchain protocol, they reflect prescribed mining resources. 

\subsection{Inferring security}

Meeting target security $S_X$ is required to produce blocks on chain $X$ in expected time $T_X$. Thus, the rate of coin issuance is tied directly to the relative difference between actual and target security. To maintain a desired block time, blockchains attempt to tune $S_X$ to match actual security $s_X$ as closely as possible. However, in PoW blockchains, $s_X$ cannot be determined from on-chain information alone. So PoW blockchain protocols must implement methods for inferring security.

\begin{mydef}
For a given blockchain, a \emph{security adjustment algorithm} (SAA) is any algorithm that adjusts its baseline security $S_X$ so that it tends toward $s_X$.
\end{mydef}

The SAA is simply an abstraction of the DAA described in Section~\ref{sec:background}. To be clear, blockchains implement DAAs, but we choose to describe them as SAAs to emphasize that they are changing the target security investment. A conventional SAA measures average block time $\bar{t}(\tau)$ over a given window and adjusts $S_X(\tau)$ in the direction of $S_X(\tau) \frac{T_X}{\bar{t}(\tau)}$. When $\bar{t}(\tau) = T_X$ we say the SAA is \emph{at rest}.

Blockchains record their security in terms of difficulty $D$. Therefore, empirical analysis requires that we express security allocation in terms of the difficulty.
Section~\ref{sec:background} describes the difficulty as the expected number of hashes required to mine a block. Thus, for chain $X \in \{A,B\}$, and when the DAA is at rest, we have that $H_X(\tau) \approx D_X(\tau) / T_X$. Finally, according to Definitions~\ref{def:chain_security} and~\ref{def:rel_sec}, 
\begin{equation}
\label{eq:infer_w}
\boldsymbol{w}(\tau) \approx \frac{1}{\hat{s}_A(\tau) + \hat{s}_B(\tau)} (\hat{s}_A(\tau), \hat{s}_B(\tau)),
\end{equation}

where $\hat{s}_X(\tau) = \frac{\sigma_X(\tau) D_X(\tau)}{T_X}$.

%% file: nash.tex
% !TEX root = main.tex

\section{Nash Equilibrium for Security Allocation}
\label{sec:nash_equilibria}

The motivating MDP in Section~\ref{sec:motivation} models a single miner, yet an essential aspect of blockchain security is competition between many miners. This nuance calls for game theory as a method to account for strategic action in anticipation of other miner's actions. Applying game theory to the hash rate allocation problem allows us to generalize the optimization problem solved by the MDP to Nash equilibria, i.e., a situation in which no party can deviate unilaterally without losing money. In fact, the equilibrium is more general in that it applies to pairs of blockchains with arbitrary difficulty and differing PoW algorithms, relative rewards, and block times. 

A relatively simple game is capable of describing the concentration point observed by the MDP. We introduce the \emph{Security Allocation Game} among two blockchains $A$ and $B$ (not necessarily sharing the same PoW algorithm), which is a one-shot game with $N$ homogeneous miners (the homogeneity assumption applies only in this section). Following conventions in the game theory literature, we distinguish an arbitrary miner $m_i$ from all the others, which are labeled $m_{\minus i}$. The miner strategy space comprises all mixed allocations among chains $A$ and $B$.

Recall from Section~\ref{sec:framework} that the security allocation across chains $A$ and $B$ for miner $m_i$ is given by $\boldsymbol{w}_i = (w_{iA}, w_{iB})$, and $\boldsymbol{w}_{\minus i} $ is similarly defined for $m_{\minus i}$. Thus, given miner $m_i$ and the group of other miners $m_{\minus i}$, the overall allocation is fully specified by $[\boldsymbol{w}_i, \boldsymbol{w}_{\minus i}]$. We assume unit aggregate security investment, which is completely allocated among the two chains, 
i.e. $|\boldsymbol{w}_i| = \frac{1}{N}$ and $|\boldsymbol{w}_{\minus i}| = \frac{N-1}{N}$. Being homogeneous, miners have the property that $(N-1) s(m_i) = s(m_{\minus i})$, i.e. each makes the same contribution to total security investment.

\makeatletter
\newcommand*\bigcdot{\mathpalette\bigcdot@{.5}}
\newcommand*\bigcdot@[2]{\mathbin{\vcenter{\hbox{\scalebox{#2}{$\m@th#1\bullet$}}}}}
\makeatother

Assuming that the SAA for each chain is at rest, the total available payoff per second is given by
\begin{equation}
\label{eq:payoff}
\boldsymbol{\pi}_i = \left(\frac{V_A}{T_A}, \frac{V_B}{T_B} \right).
\end{equation}
Now define 
\begin{equation}
\boldsymbol{u}_i = \left( \frac{w_{iA}}{w_{iA} + w_{\minus iA}}, \frac{\frac{1}{N} - w_{iA}}{1 - w_{iA} - w_{\minus iA}} \right),
\end{equation}
which is miner $m_i$'s share of the reward on each chain. The payoff for $m_i$ is equal to $\boldsymbol{\pi}_i^\textsc{t}  \boldsymbol{u}_i$. Payoff has one term per chain and reflects the fact that reward is distributed to miners (in expectation)
proportionally to the security they allocate to each chain. 
We search for a pure-strategy Nash equilibrium that leverages the payoff function in Eq.~\ref{eq:payoff} and the miner homogeneity assumption. The existence of a symmetric pure strategy equilibrium is not remarkable, but it is instructive to show that such an equilibrium matches the main equilibrium discovered by Kwon et al.~\cite{Kwon:2019}.

\smallskip
\begin{mythm}
\label{thm:nash_equil}
The following allocation is a symmetric equilibrium for the Security Allocation Game: 
\[
[\boldsymbol{w}^*_i, \boldsymbol{w}^*_{\minus i}] = \left[ \frac{1}{N} (c, 1-c), \frac{n}{N} (c, 1-c) \right], 
\]
where $n = N-1$ and $c = \frac{T_B R}{T_B R - T_A R + T_A}$. When $T_A = T_B$ the equilibrium simplifies to $c = R$. (Proof in Appendix~\ref{sec:proofs}.)
\end{mythm}

The equilibrium specified by Theorem~\ref{thm:nash_equil} coincides with the concentration point identified in Figure~\ref{fig:mdp_a6}. This result tells us that a relatively simple game theoretical model explains the behavior observed in the optimal solution to a specific hash rate allocation problem, but with greater generality. 
However, the game theoretical approach also carries significant limitations. 
First, it relies on the homogeneity of hash power among miners. 
Second, it assumes that all miners have the same utility (optimizing Eq.~\ref{eq:payoff}), which is unrealistic because miners face variable costs and they may accept losses to promote a chain of their liking.
Third, our simple game assumes that miners have no outside options, which exist in the real world by abstaining, mining on a third chain, or selling excess mining capacity.
Fourth, the game does not consider higher moments of the payoff distribution (beyond expected value): miner risk appetite might result in different adjustments to obtain their individual objective function.
Fifth, the approach is not exhaustive. It is difficult to completely eliminate the possibility of other equilibria that might arise asymmetrically or in mixed strategies.

Kwon et al.~\cite{Kwon:2019} also developed a game theoretical model, which identifies the same symmetric Nash equilibrium and a number of others. However, that model suffers from the same limitations listed above and others as well. First, that model defines several game players: \emph{loyal, automatic, stick} and \emph{fickle} miners. Such a composition of players is problematic: some are implausible (for example, fickle miners switch chains \emph{only} when the difficulty adjusts) and their relative quantities are unobservable. Second, that model describes equilibrium strategies for fickle miners by making assumptions about the resources and behavior of the loyal and stick miners. As a result, the utility function for fickle miners incorporates the characteristics of loyal and stick miners, \emph{which cannot be known a priori}. This assumption makes it even less plausible that miners would all follow the same utility function. Note that our model assumes that best-response miners can switch chains instantaneously. We feel that this is the most plausible game player; and we show in Section~\ref{sec:eval} that it is sufficient to explain much of miner behavior in practice.

In the next section, we introduce an approach that captures the uniqueness of the equilibrium described in  Theorem~\ref{thm:nash_equil}. We use the technique to show that this equilibrium is unique under much weaker assumptions, eliminating the list of limitations above.

%% file: na.tex
% !TEX root = main.tex

\begin{figure}[t]
\centerline{\includegraphics[trim=0 10 0 0,clip,width=0.55\linewidth]{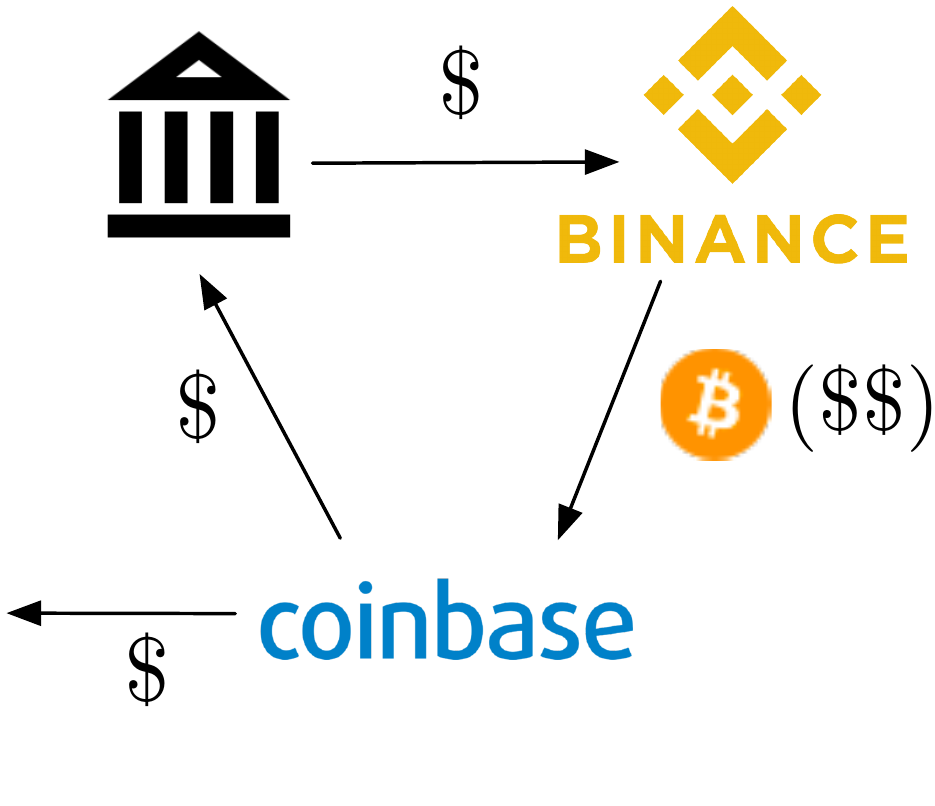}}
\vspace*{-5mm}
\caption{\emph{Exploiting conventional (two-point) arbitrage in currency exchange. An investor moves fiat currency from a bank to the Binance exchange, where she trades the fiat for Bitcoin. She then transfers the Bitcoin to the Coinbase exchange where she is able to sell it for a greater amount of fiat currency than she began with.}}
\vspace{-1em}
\label{fig:arb1}
\end{figure}

\section{Arbitrage Conditions at Equilibrium}
\label{sec:arb_conditions}

The game theoretical equilibrium of Section~\ref{sec:nash_equilibria} is important because it explains the behavior of a group of strategic miners who understand how mining profitability changes in a competitive environment. In the case where all miners achieve this level of sophistication, and subject to the assumptions of the section, the equilibrium of Theorem~\ref{thm:nash_equil} will be achieved. However, it is unlikely that all miners currently are strategic and implausible that all assumptions are met. Given these limitations, we seek to understand how security allocation is affected when most of the assumptions in Section~\ref{sec:nash_equilibria} are relaxed. 

The finance literature has studied \emph{no arbitrage} (NA) conditions, weak conditions that guarantee profitability. Informally, arbitrage occurs when profit is made at zero cost. NA theory posits that agents will change their behavior to exploit arbitrage opportunities when they exist, and will maintain their behavior (forming an equilibrium) when they do not. Figure~\ref{fig:arb1} depicts conventional two-point arbitrage in the context of currency exchange. In this example, an investor sees an opportunity to capitalize on the difference in trade price of Bitcoin on two different exchanges. The existence of arbitrage creates strong incentive for investors to exploit the opportunity until they reach a point of no arbitrage.

We prove Theorems~\ref{thm:equil} and~\ref{thm:na_unique} and Corollary~\ref{cor:na_unique}, which together imply that: \1 there exists a single security allocation $\boldsymbol{w}_{\texttt{eq}}$ that achieves no arbitrage; \2 at every other allocation, it is possible to exploit arbitrage by rebalancing in the direction of $\boldsymbol{w}_{\texttt{eq}}$; and \3 as long as miners maintain constant security (i.e., hash rate) across chains, \emph{every} rebalancing that exploits arbitrage will move the allocation in the direction of $\boldsymbol{w}_{\texttt{eq}}$. Based on these results, \emph{our key finding is that a miner allocating hash rate off equilibrium (be it accidental or intentional) will not tend to move the equilibrium because his bias toward one chain will be offset by another miner exploiting the resulting arbitrage opportunity.} Therefore, to boost the proportion of hash rate on a given chain, one must move the market price of the chain's coin, not donate hash power.

\subsection{Derivative Markets} 

In the study of finance, we are often concerned with the payoff of a certain \emph{portfolio} of \emph{financial securities} at a future date. In the simplest model, agent $m_i$ purchases \emph{contingent claims} on securities $A$ and $B$ in quantities $\boldsymbol{c}_i(\tau_1) = (c_{iA}(\tau_1), c_{iB}(\tau_1))$ at time $\tau_1$ using an initial \emph{endowment} $e_i(\tau_1)$. The endowment constitutes the resources available to the agent for purchasing contingent claims. A contingent claim is any sort of derivative contract on the security whose payout depends on a future \emph{outcome}, such as an option or futures contract. Contingent claims carry purchase prices $\boldsymbol{p}(\tau_1) = (p_A(\tau_1), p_B(\tau_1))$ at $\tau_1$. Naturally, the \emph{portfolio price}, $(\boldsymbol{c}_i(\tau_1))^\textsc{t}  \boldsymbol{p}(\tau_1)$, must not exceed endowment $e_i(\tau_1)$, which is the agent's budget constraint. Contingent claims can be sold at time $\tau_2$ for payoff $\Pi(\tau_2)$, where $\Pi(\tau_2)$ is a matrix with columns corresponding to portfolio components, rows corresponding to individual states, and where each matrix entry corresponds to an outcome, claim pair. The agent seeks to maximize aggregate payoff, $\sum_{X \in \{A,B\}} (\Pi(\tau_2) c_i(\tau_1))_X$.

\subsubsection{Notation}

In the remainder of this section, we occasionally drop the time argument $\tau$ where it can be understood from context, but we reintroduce it in places where time should be emphasized. Also, for ease of exposition, we use \emph{Hadamard notation} for component-wise multiplication and division of vectors $\boldmath{u}$ and $\boldmath{v}$: 
\[
\boldsymbol{u} \odot \boldsymbol{v} = (u_A v_A, u_B v_B)
\mbox{~~~and~~~}\boldsymbol{u} \oslash \boldsymbol{v} = (u_A / v_A, u_B / v_B).
\]

\subsection{Blockchain Security Market}

We define the \emph{Blockchain Security Market} for agent $m_i$, a miner, as follows. Endowment $e_i(\tau_1)$ is equal to $s(m_i)$, or the fraction of all fiat currency devoted to security across chains $A$ and $B$ at time $\tau_1$ that belongs to $m_i$. Accordingly, $e_i(\tau_1)$ is also a scalar multiple of $H(m_i)$, the number of hashes that $m_i$ is capable of producing per second. We assume that $e_i(\tau_1)$ remains fixed over time so that the miner consistently operates with the same hash rate. 

Each portfolio, $\boldsymbol{c}_i(\tau_1)$, represents a contingent claim on future coinbase from chains $A$ and $B$, respectively, between times $\tau_1$ and $\tau_2$, where we assume that $\tau_2 - \tau_1 = 1$ second. Price vector $\boldsymbol{p}(\tau_1) = (S_A(\tau_1), S_B(\tau_1))$ is equal to the cost of \emph{purchasing} 1 second worth of expected reward for mining on chains $A$ and $B$. Claim vector $\boldsymbol{c}_i(\tau_1)$ for miner $i$ is expressed as a fraction of $\boldsymbol{p}(\tau_1)$, but such that the fraction can exceed 1; i.e., it is possible to purchase more than a single claim each second (which would tend to generate blocks faster than the blockchain's target rate). We consider only one state at time $\tau_2$, having payoff vector \vspace{-.5ex}
\[
\boldsymbol{\pi}(\tau_2) = \left( \frac{V_A(\tau_2)}{t_A(\tau_2)}, \frac{V_B(\tau_2)}{t_B(\tau_2)} \right),
\] 
which is the total expected fiat value for each chain's block reward during the 1 second time period and is equal to Eq.~\ref{eq:payoff} when SAAs for chains $A$ and $B$ are at rest. 

In this definition, there exists no contingency because there is only one possible state at time $\tau_2$. As such, it is possible to guarantee payoff $ \boldsymbol{c}_i(\tau_1)^\textsc{t}  \boldsymbol{\pi}(\tau_2)$ at $\tau_2$. Finally, we redefine the \emph{security allocation} for miner $m_i$ at time $\tau_1$, in the context of the blockchain market, by \vspace{-.5em}
\begin{equation}
\label{eq:allocation}
\boldsymbol{w}_i(\tau_1) = \frac{1}{e_i(\tau_1)} \boldsymbol{c}_i(\tau_1) \odot \boldsymbol{p}(\tau_1).
\end{equation} 
This allocation corresponds to the fraction of the miner's total security investment devoted to each chain. Throughout, we assume that $|\boldsymbol{w}_i| = 1$, in other words, the miner allocates his resources entirely among the two chains.

\subsubsection{Portfolio Rebalancing}
We imagine that each miner holds initial claim $\boldsymbol{c}_i(\tau)$ but wishes to \emph{rebalance} to a new claim $\boldsymbol{c}_i(\tau')$ at some future time $\tau'$ with the hope of achieving a higher payoff. Prices at time $\tau'$, $\boldsymbol{p}(\tau') = (S_A(\tau'), S_B(\tau'))$, correspond to the target security investment (fiat value per second) required to mine a block on each chain in the desired expected time. For the miner to rebalance his claims, he must \emph{sell short} his claim on one chain to increase his claim on another. Thus,  to enforce the notion of scarcity in security investment (and ultimately hash rate), we stipulate that $c_{iX}(\tau') p_X(\tau') \geq c_{iX}(\tau) p_X(\tau)$, $X \in \{A, B\}$. 
This stipulation implies that, on any given chain, the miner cannot sell short a claim at price $\boldsymbol{p}(\tau')$ with total fiat value exceeding what he purchased at time $\tau$. 

\subsubsection{Properties of security allocations}

\begin{figure}[t]
\centerline{\includegraphics[trim=0 10 0 0,clip,width=0.65\linewidth]{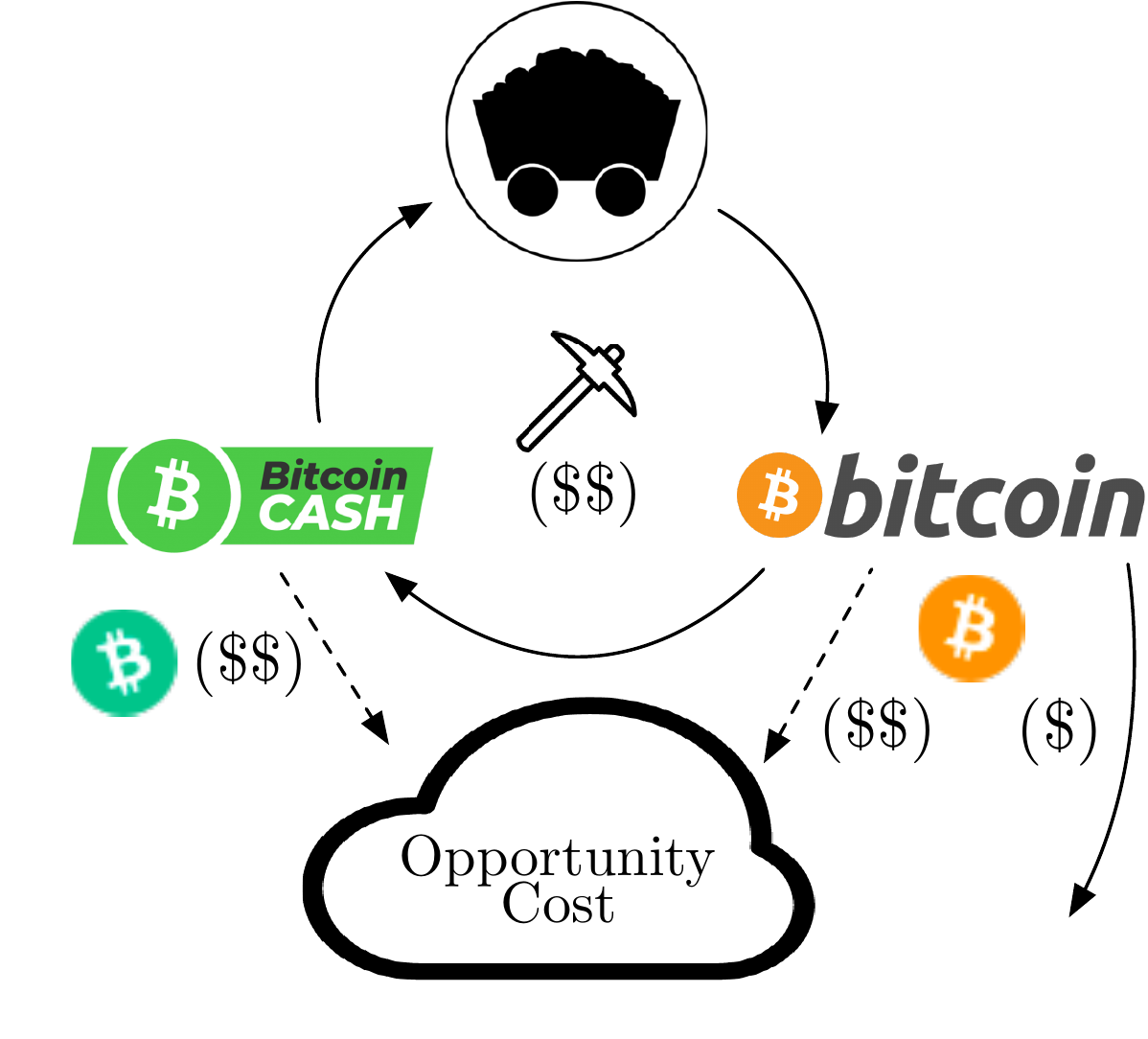}}
\vspace*{-2mm}
\caption{\emph{Exploiting arbitrage between chains sharing the same PoW algorithm in the Blockchain Security Market. Hashes, having opportunity cost $(\$\$)$, are generated at a constant rate by the miner. They can be traded off between the Bitcoin Cash and Bitcoin blockchains. In this example, shifting hashes to the Bitcoin blockchain will mine Bitcoin having value $(\$)$ in excess of the opportunity cost. This results in a Bitcoin payoff  of  value  $(\$)$.}}
\label{fig:arb2}
\end{figure}

We are primarily interested in the overall effect of miner behavior on the equilibrium of Theorem~\ref{thm:nash_equil}. Therefore, going forward, we consider only \emph{aggregate} allocations. 

\begin{mydef}
\label{def:agg_alloc}
The \emph{aggregate} claim $\boldsymbol{c}$ and endowment $e$ across multiple miners are given by $\sum_i \boldsymbol{c}_i$ and $\sum_i e_i$, respectively. Aggregate security allocation is $\boldsymbol{w} = \boldsymbol{c} \odot \boldsymbol{p} / e$. 
\end{mydef}

Note that Definition~\ref{def:agg_alloc} provides a reinterpretation of security allocation $w$ from  Definition~\ref{def:rel_sec} in terms of portfolio price and claims.
\noindent The following definitions are useful to us for reasoning about changes in allocation.

\begin{mydef}
\label{def:distance}
The \emph{distance} between two allocations $\boldsymbol{w}_1$ and $\boldsymbol{w}_2$ is given by the L1-norm of their difference: $|\boldsymbol{w}_1 - \boldsymbol{w}_2|$.
\end{mydef}

\begin{mydef}
An \emph{allocation rebalancing} is an allocation $\Delta \boldsymbol{w}$ intended to update existing allocation $\boldsymbol{w}$ to $\boldsymbol{w}' = \boldsymbol{w} + \Delta \boldsymbol{w}$. We say that a rebalancing is \emph{symmetric} when $\Delta \boldsymbol{w} = (\epsilon, -\epsilon)$ for some $\epsilon \in \mathbb{R}$.
\end{mydef}

We are primarily interested in symmetric allocation rebalancings because they maintain constant aggregate resources across both chains.

\subsubsection{Portfolio pricing}

Typically, security prices emerge when buyers and sellers come to an agreement on an exchange price but the blockchain security market is unique in that prices are set algorithmically by the SAA. At time $\tau$ on blockchain $X$, the SAA responds to a difference between the baseline security $S_X(\tau)$ and the inferred actual security $s_X(\tau)$ by moving the value of the former closer to the value of the latter. And because $p_X(\tau) = S_X(\tau)$, the action of the SAA has the effect of changing the portfolio price. The following proposition shows how to determine portfolio price when the SAA is at rest, i.e. $S_X(\tau) = s_X(\tau)$.

\bigskip
\begin{myprop}
\label{thm:portfolio_price}
For any fixed allocation $\boldsymbol{w}$, after the SAAs on chains $A$ and $B$ come to rest, the  portfolio pricing vector will be $\boldsymbol{p} = e \boldsymbol{w}$. (Proof in Appendix~\ref{sec:proofs}.)
\end{myprop}

\subsubsection{Arbitrage}

\begin{figure}[t]
\centerline{\includegraphics[trim=0 10 0 0,clip,width=0.55\linewidth]{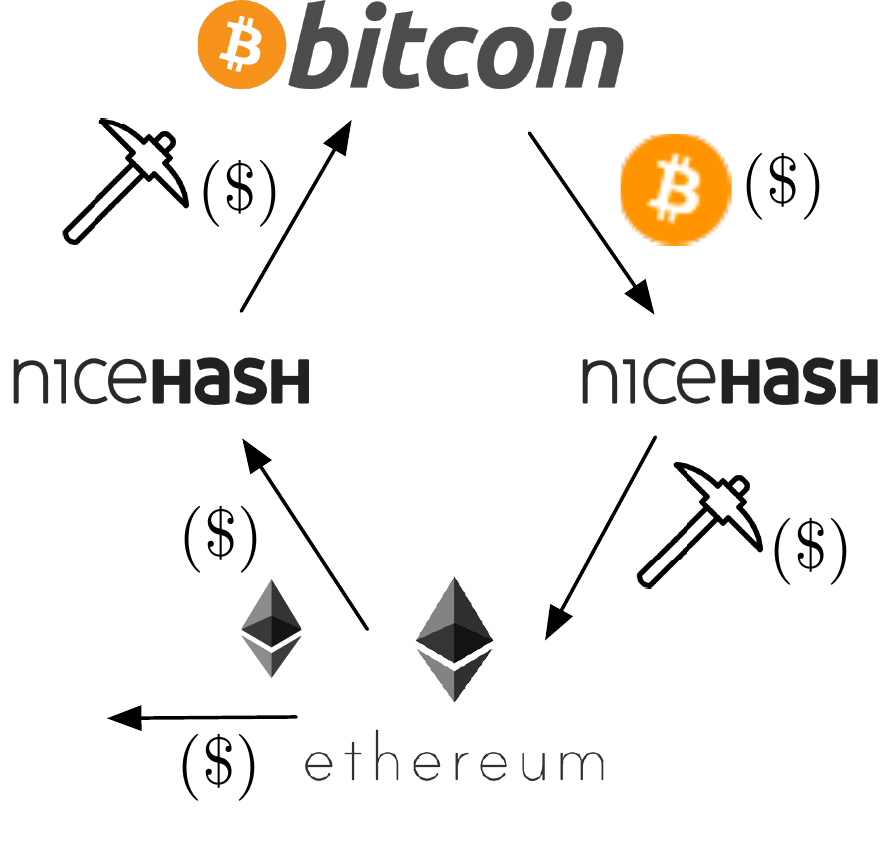}}
\vspace*{-2mm}
\caption{\emph{Exploiting arbitrage between chains having different PoW algorithms in the Blockchain Security Market. Proceeding clockwise from the upper left,  SHA256 hash rate having fiat value $(\$)$ is used to mine Bitcoin for zero marginal profit. The Bitcoin is then transferred back to NiceHash and used to purchase Dagger-Hashimoto hash rate at cost $(\$)$. This hash rate is applied to the Ethereum blockchain, which yields ether having fiat value $(\$\$)$, and which is $(\$)$ greater than the fiat value of the original SHA256 hash rate.}}
\label{fig:arb3}
\end{figure}

An \emph{arbitrage opportunity} is formally defined as the circumstance where $\boldsymbol{\pi}^\textsc{t} \Delta \boldsymbol{w} \geq 0$ and $\Delta \boldsymbol{w}^\textsc{t}  \boldsymbol{p} \leq 0$, with at least one strict inequality~\cite{LeRoy:2014}. Less formally, arbitrage is possible any time it is possible to guarantee future payoff at zero cost. We expect that a miner will seek to rebalance his claim to exploit the higher payoff in this circumstance. Figure~\ref{fig:arb2} shows how a miner can rebalance his claim (i.e., his hash rate) between two blockchains sharing the same PoW algorithm  to increase his profit. In this case, the opportunity cost of mining is considered a sunk cost, and arbitrage is captured by shifting hash rate to Bitcoin, which is the more profitable chain. In contrast, Figure~\ref{fig:arb3} illustrates how arbitrage can be exploited among blockchains that generally employ different PoW algorithms by an agent owning no hash rate. The agent trades fiat for hash rate in a market such as NiceHash~\cite{nicehash}, and then distributes those hashes among blockchains. In this example, the fiat reward per unit of cost to secure Ethereum is greater than in Bitcoin. Therefore, it is possible to trade hash rate on Bitcoin for hash rate on Ethereum to boost profits. 

We next prove that the equilibrium allocation defined in Theorem~\ref{thm:nash_equil} is a point of no arbitrage. It is a point where there exists no financial incentive for miners to rebalance their portfolio of security allocations. 

\smallskip
\begin{mythm}
\label{thm:equil}
Assume any choice of SAA for chains $A$ and $B$ (not necessarily the same). When the relative reward $R$ is stable, there exists no arbitrage at the following allocation 
\begin{equation}
\label{eq:equil}
\hspace{-.75em}\boldsymbol{w}_{\texttt{eq}} = \left( \frac{T_B R}{T_B R - T_A R + T_A}, \frac{T_A (1 - R)}{T_B R - T_A R + T_A} \right),
\end{equation}
which simplifies to 
\begin{equation}
\boldsymbol{w}_{\texttt{eq}} = (R, 1 - R),
\end{equation}
if $T_A = T_B$. (Proof in Appendix~\ref{sec:proofs}.)
\end{mythm}

Notice that the sole requirement for maintaining the equilibrium described by Theorem~\ref{thm:equil} is for each miner to actively update his allocation so as to maximize profit. Consequently, the no arbitrage equilibrium is more plausible in practice than a Nash equilibrium because it can be achieved without a complex utility function and without directly contemplating the actions of other miners.

\subsubsection{Uniqueness}

Now we  establish the uniqueness of the equilibrium defined by Eq.~\ref{eq:equil} among all potential points of no arbitrage, which further motivates its formation in practice. To that end, we begin by proving a lemma that shows portfolio cost remains unchanged by any symmetric rebalancing.

\begin{mylem}
\label{eq:sym_rebalance}
For initial allocation $\boldsymbol{w}$ and price $\boldsymbol{p}$, the claims associated with a symmetric rebalancing  $\Delta \boldsymbol{w}$ are given by $\Delta \boldsymbol{c} = \Delta \boldsymbol{w} \oslash \boldsymbol{w}$ and it is always the case that $\Delta \boldsymbol{c}^\textsc{t}   \boldsymbol{p} = 0$. (Proof in Appendix~\ref{sec:proofs}.)
\end{mylem}

In the following theorem, we establish that there exists opportunity for arbitrage at \emph{any} allocation that is distinct from the equilibrium defined in Eq.~\ref{eq:equil}. Moreover, we show that the arbitrage can be exploited with a symmetric rebalancing that moves the allocation closer to the equilibrium. 

\smallskip
\begin{mythm}
\label{thm:na_unique}
For any allocation $\boldsymbol{w} \neq \boldsymbol{w}_{\texttt{eq}}$, there exists a symmetric allocation rebalancing $\Delta \boldsymbol{w}$, such that $|(\boldsymbol{w} + \Delta \boldsymbol{w}) - \boldsymbol{w}_{\texttt{eq}}| \leq |\boldsymbol{w} - \boldsymbol{w}_{\texttt{eq}}|$, which has price zero and strictly positive payoff. (Proof in Appendix~\ref{sec:proofs}.)
\end{mythm}

We close this section by arguing that, for allocations not at the equilibrium defined by Eq.~\ref{eq:equil}, every symmetric rebalancing that moves the allocation away from the equilibrium can only reduce the miner's payoff. This result is significant because, along with Theorem~\ref{thm:na_unique}, it establishes that the equilibrium of Theorem~\ref{thm:equil} is an \emph{attractor}, meaning that off-equilibrium allocations will tend to rebalanced toward it.

\smallskip
\begin{mycor}
\label{cor:na_unique}
For allocation $\boldsymbol{w} \neq \boldsymbol{w}_{\texttt{eq}}$, any symmetric rebalancing allocation $\Delta \boldsymbol{w}$ such that $|(\boldsymbol{w} + \Delta \boldsymbol{w}) - \boldsymbol{w}_{\texttt{eq}}| > |\boldsymbol{w} - \boldsymbol{w}_{\texttt{eq}}|$ has price zero will result in strictly negative payoff. (Proof in Appendix~\ref{sec:proofs}.)
\end{mycor}

The results of this section prove that the only allocation with no arbitrage is at $\boldsymbol{w}_{\texttt{eq}}$ and that exploiting arbitrage at any other allocation will move it closer to $\boldsymbol{w}_{\texttt{eq}}$. This result has important implications for blockchain security and governance. A major conclusion of Kwon et al.~\cite{Kwon:2019} is that, subject to the assumptions of their Nash equilibrium, miners loyal to one chain (which they call \emph{stick}) will mine alone if their allocation exceeds $\boldsymbol{w}_{\texttt{eq}}$. We deepen this result by showing it holds under the much weaker assumptions of NA theory. \emph{As long as most miners act to exploit arbitrage, allocation will always return to equilibrium.} And our results further show that loyal miners confer no marginal improvement in security to a chain for any hash rate that they contribute below $\boldsymbol{w}_{\texttt{eq}}$ --- if their hash rate was absent, then it would be replaced by ordinary miners exploiting arbitrage.

%% file: evaluation.tex
% !TEX root = main.tex

\section{Evaluation}
\label{sec:eval}

In this section, we demonstrate empirically the formation of the security allocation equilibrium described variously in Sections~\ref{sec:motivation}, \ref{sec:nash_equilibria}, and \ref{sec:arb_conditions}. Our theory is overwhelmingly supported by data at hourly granularity, with much lower error results than previous work~\cite{Kwon:2019} that used less granular data. Moreover, we illuminate security allocation relationships between blockchains previously believed to be unrelated.

Recall that the \emph{actual} security allocation between two blockchains $A$ and $B$ is given by $\boldsymbol{w}$ (see Definition~\ref{def:rel_sec}). In plain terms, a certain amount of hash rate is applied to both chains cumulatively and this hash rate has a fiat value (as determined by its trade price $\sigma$ in a marketplace like NiceHash~\cite{nicehash}). Vector $\boldsymbol{w}$ captures the relative fiat value devoted to security on each chain. 
Below, we show that the equilibrium point $\boldsymbol{w}_e$, described by Theorem~\ref{thm:equil}, closely matches the actual allocation $\boldsymbol{w}$ for historical data. 

\subsection{Data collection and preprocessing}

We collected historical data for several of the largest PoW blockchains by market capitalization including Bitcoin (BTC), Bitcoin Cash (BCH), Ethereum (ETH), Ethereum Classic (ETC), and Litecoin (LTC). Included in the datasets were hourly fiat / coin exchange prices from the Bitfinex exchange~\cite{bitfinex} for dates prior to November 15, 2018 and from the Coinbase exchange~\cite{coinbase} for dates after. Data from the Bitstamp~\cite{bitstamp} exchange were used for BCH only for the seven days following a contentious hard fork on November 15, 2018. We used publicly available Blockchain data  in the Google BigQuery database~\cite{bigquery}. We adjusted Blockchain constants such as target block time and block reward over time to match historical values. We gathered  hash price data from NiceHash~\cite{nicehash} for dates on or after October 10, 2019. We downloaded the hash price order book from NiceHash every 10 minutes, and used the mean price from those orders as the spot price. Our results are not significantly different when either median or best prices are used instead. 

\subsubsection{Estimating security allocation $\boldsymbol{w}$}
We calculated the security allocation $\boldsymbol{w}$ between pairs of blockchains  using Eq.~\ref{eq:infer_w}. This required calculating $\hat{s}$, the estimated security investment for a given blockchain, whose major components are hash rate $H$ and hash price $\sigma$. From the blockchain data, we were able to extract the \emph{nominal} hash rate $H'(\tau)$ at time $\tau$ from its difficulty $D(\tau)$. For ETH and ETC, nominal hash rate is simply $H'(\tau) = D(\tau) / T$, where $T$ is the target block time. For BTC, BCH, and LTC it is also necessary to multiply by \emph{pool difficulty} (\url{https://en.bitcoin.it/wiki/Difficulty}), so that $H'(\tau) = 2^{32} D(\tau) / T$. $H'$ was a sufficiently smooth estimator for all blockchains except BTC and LTC, which update their difficulty only once every 2016 blocks. 
For these two chains, we estimated the hash rate at time $\tau$, $\hat{H}(\tau)$, by adjusting a rolling nominal hash rate with a rolling correction term based on observed block times. In particular
\begin{equation}
\hat{H}(\tau) = \frac{\texttt{ewma}(H'(\tau)) }{\texttt{ewma}(t(\tau))} T,
\end{equation}
where $t(\tau)$ is the actual block time for the given chain and $\texttt{ewma}$ denotes the exponentially weighted moving average until time $\tau$ with 96-hour half-life. The choice of a 96-hour half-life is justified by Figure~\ref{fig:me-within-epoch} (see Appendix~\ref{sec:suppl_figs}), which shows that it is capable of correcting a systematic bias in nominal hash rate.
Thus, for input to Eq.~\ref{eq:infer_w} we use \vspace{-.5em}
\begin{equation}
\hat{s}(\tau) = 
\texttt{ewma}(\sigma(\tau)) \frac{\texttt{ewma}(H'(\tau)) }{\texttt{ewma}(t(\tau))} T.
\end{equation}

\definecolor{red}{RGB}{212,133,144	}
\definecolor{blue}{RGB}{100,149,236}
\begin{figure}[t]
\includegraphics[trim=0 15 0 10,clip,width=\linewidth]{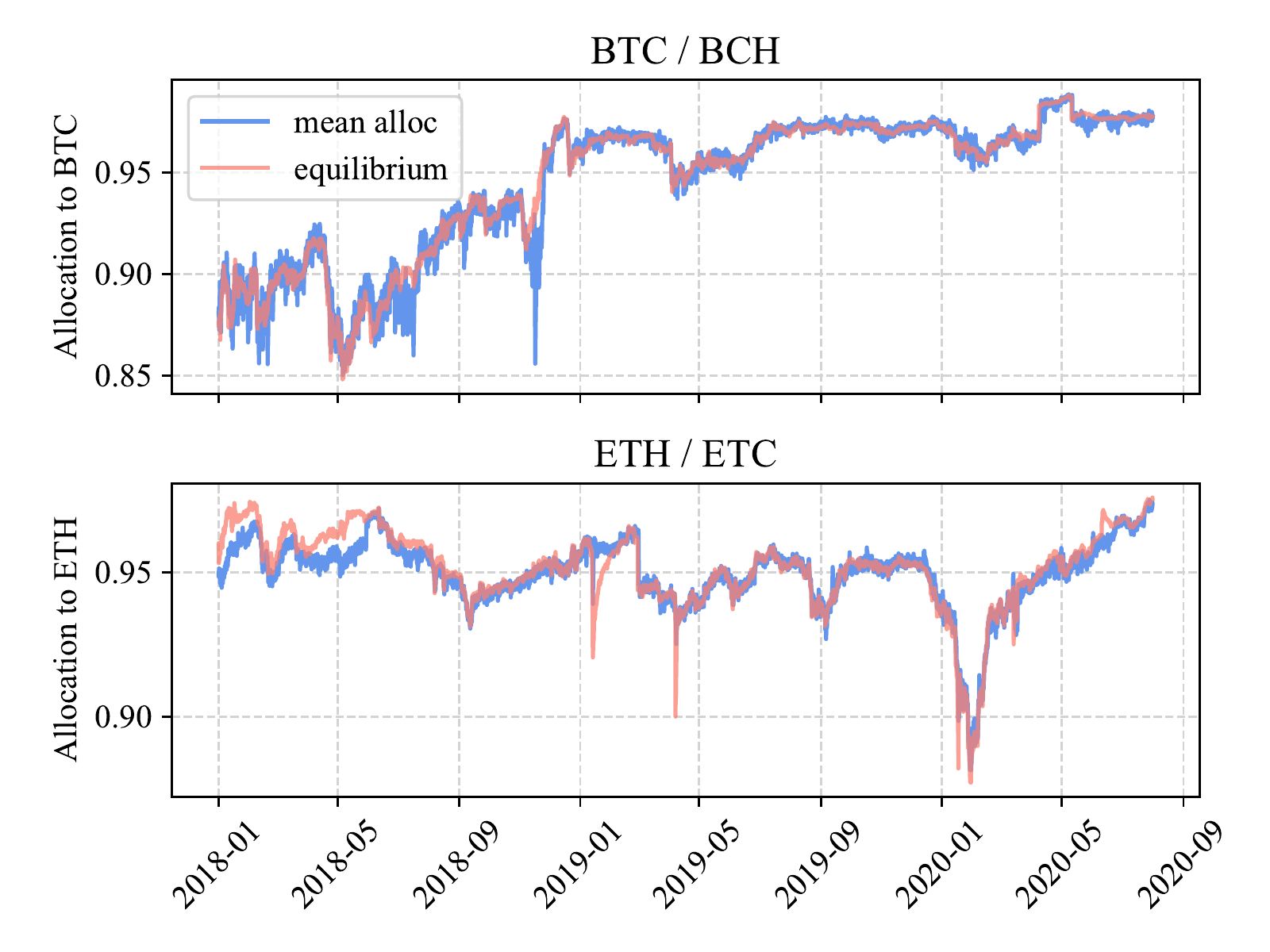}
\caption{\emph{Actual hash rate allocation ({\color{blue}blue}) between two 
cryptocurrencies having the \textbf{same PoW algorithm} juxtaposed with the equilibrium allocation ({\color{red}red}). The plots show strong agreement between the actual allocation and the allocation predicted by the equilibrium, the latter of which is based entirely on expected block times and coinbase values. Spikes in ETH / ETC are due to large transaction fees miners couldn't anticipate.}}
\label{fig:single_pow_eq}
\end{figure}

\begin{figure}[t]\vspace{-.5cm}
\centerline{\includegraphics[trim=0 15 0 10,clip,width=\linewidth]{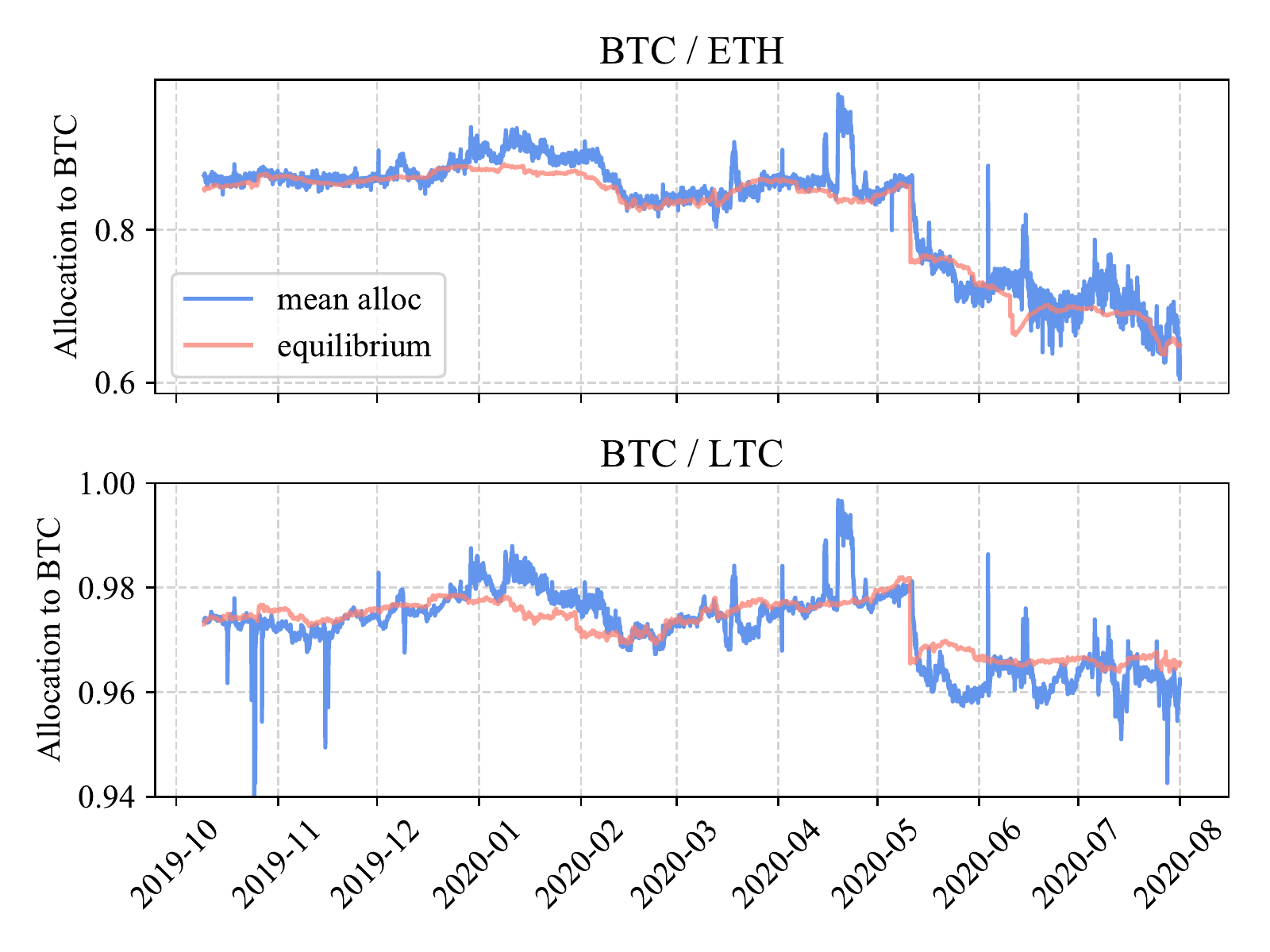}}
\caption{\emph{Actual hash rate allocation ({\color{blue}blue}) between two pairs of cryptocurrencies using \textbf{different PoW algorithms} juxtaposed with the equilibrium allocation ({\color{red}red}). Agreement with the equilibrium is strong, albeit with significant bias during January and after May, 2020. Spikes in actual allocation are an artifact of sudden changes in hash price in the NiceHash marketplace. The   BTC / LTC plot is trimmed and excludes larger spikes, the lowest spike dips to nearly 0.92.}}
\label{fig:multi_pow_eq}
\end{figure}

\subsubsection{Estimating equilibrium between chains $\boldsymbol{w}_{\texttt{eq}}$}

\begin{figure}
\centering
\begin{tikzpicture}
	\node [inner sep=0] (A) {\includegraphics[width=0.9\linewidth]{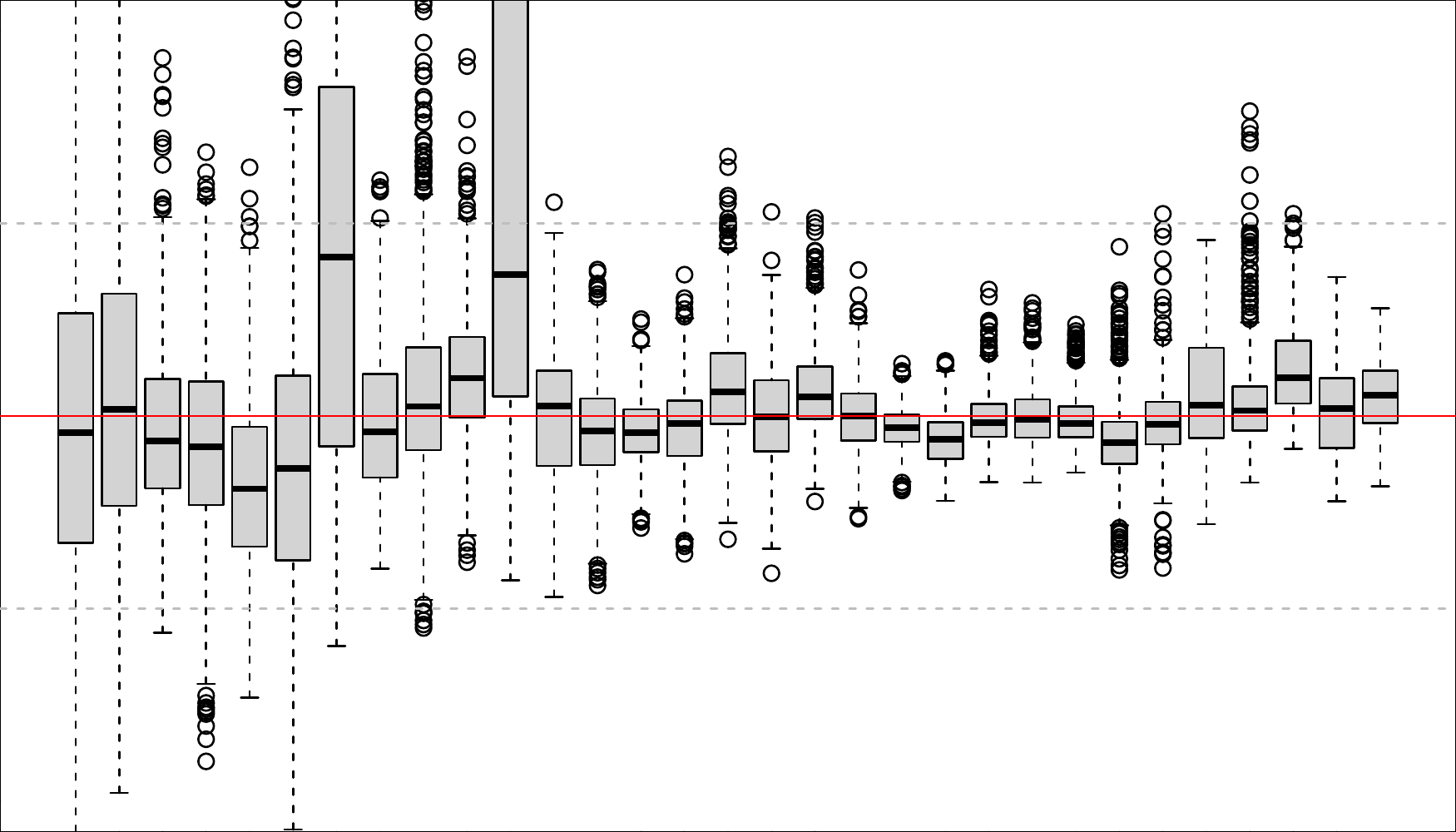}};
	\scriptsize
	\draw (A.west) node [left,red] {$0$};
	\draw [white] (A.west) -- node [pos=.48,left,black!50] {$-0.01$} (A.south west);
	\draw [white] (A.west) -- node [pos=.48,left,black!50] {$0.01$} (A.north west);

	\foreach \x in {0,12}
		\draw [x=2.39mm,anchor=base] (A.south west)++(\x,-2ex)++(1.6,0)
			node (J\x) {\,J} ++(1,0) node {F} ++(1,0) node {M} ++(1,0) node {A} ++(1,0) 
			node {M} ++(1,0) node {J} ++(1,0) node {J} ++(1,0) node {A} ++(1,0) 
			node {S} ++(1,0) node {O} ++(1,0) node {N} ++(1,0) node (D\x) {D\vphantom{J}} ++(1,0);
	\foreach \x in {24}
		\draw [x=2.39mm,anchor=base] (A.south west)++(\x,-2ex)++(1.6,0)
			node (J\x) {\,J} ++(1,0) node {F} ++(1,0) node {M} ++(1,0) node {A} ++(1,0) 
			node {M} ++(1,0) node {J} ++(1,0) node (JX) {J} ++(1,0); % node {A} ++(1,0) ;
			
	\draw (J0.south west)-- node [below] {\textbf{2018}} (D0.south east);
	\draw (J12.south)-- node [below] {\textbf{2019}} (D12.south east);
	\draw (J24.south)-- node [below] {\textbf{2020}} (JX.south east);
\end{tikzpicture}

\caption{\emph{Evolution of the BTC / BCH prediction error over time. 
Monthly distributions of hourly differences between equilibrium and actual hash rate allocation (in allocation units). 
Positive values mean that, from market prices, our theory predicts a higher allocation of hash power to BTC than the actual value.}}
\vspace{-0.5em}
\label{fig:me-months-btc-bch}
\end{figure}
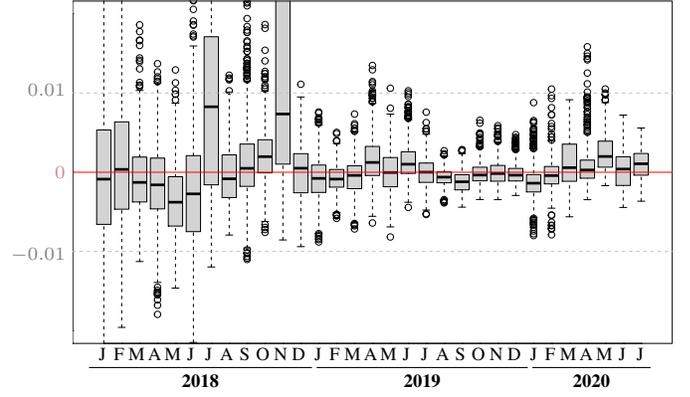

We calculated the equilibrium allocation between two blockchains $\boldsymbol{w}_{\texttt{eq}}$   using Eq.~\ref{eq:equil}. The only variable quantity in that equation is $R$, which is a function of the fiat value of rewards, $V(\tau)$, paid out each block on either chain. At time $\tau$, fiat reward itself was calculated as the product of coinbase reward plus fees in the native currency, $k(\tau)$, and the fiat exchange rate, $P(\tau)$. For $P(\tau)$ we used the average of the high and low prices for each hour. Reward $k(\tau)$ varied only with transaction fees, which were highly variable for all blockchains. We smoothed fee values for each chain using an exponentially weighted moving average with a half-life of 96 hours. Smoothing fees is justified by the fact that miners cannot always redirect their mining resources in time to capitalize on an unusually high transaction fee, so they are more likely to assume average rather than instantaneous fees. Generally, multiple blocks arrived per hour, so to align coinbase reward with coin price (measured hourly), we used the average reward per hour.

\subsection{Historical Convergence to Equilibrium}

Figure~\ref{fig:single_pow_eq} plots the actual security allocation $\boldsymbol{w}$ in blue for  pairs of blockchains BTC / BCH (utilizing the SHA256 algorithm) and ETH / ETC (utilizing the DaggerHashimoto algorithm), with one pair per facet, along with the equilibrium allocation $\boldsymbol{w}_{\texttt{eq}}$, which is plotted in red. Table~\ref{tab:gof-indicators} also shows several error metrics for each pair, broken down by year. Overall agreement between the red and blue curves was excellent in both facets, which indicates convergence to the equilibrium defined by Theorem~\ref{thm:equil}. The most notable deviation from equilibrium in the BTC / BCH plot occurs during a contentious hard fork on the BCH chain (which created Bitcoin Satoshi Vision or BSV; see \url{https://en.wikipedia.org/wiki/Bitcoin_Cash}). The fork was responsible for draining hash rate from the BCH chain and also resulted in roughly a week-long halt to nearly all fiat exchange of the BCH coin. Both these factors likely contributed to the disruption in the equilibrium. Figure~\ref{fig:me-months-btc-bch} shows the trend in prediction error for the BTC / BCH pair over time. Agreement between actual and equilibrium allocations has tightened considerably since 2019 with bias becoming particularly low. 

\begin{table}
{\relsize{-1}
\begin{tabular}{p{2em}ll rrrr}
\toprule
&&
\multicolumn{1}{c}{\bf RMSE} & 
\multicolumn{1}{c}{\bf MAE} & 
\multicolumn{1}{c}{\bf ME} & 
\multicolumn{1}{c}{\bf PSNR} \\
\midrule
\midrule
\textbf{2018} 
	& BTC/BCH &
	$0.0091$&$0.0058$&$-0.0012$&$40.8669$ \\
	& ETH/ETC &
	$0.0059$&$0.0042$&$-0.0042$&$44.5371$ \\
\midrule
\textbf{2019} 
	& BTC/BCH &
	$0.0021$&$0.0016$&$0.0005$&$53.5354$ \\
	& ETH/ETC &
	$0.0040$&$0.0019$&$0.0009$&$48.0315$ \\
	& BTC/ETH$\ast$ & 
	$0.0096$&$0.0071$&$0.0051$&$40.3168$ \\
	& BTC/LTC$\ast$ & 
	$0.0030$&$0.0020$&$-0.0014$&$50.3599$ \\
\midrule
\bf 2020 
    & BTC/BCH &
	$0.0026$&$0.0019$&$-0.0003$&$51.5503$ \\
\bf (thru	
    & ETH/ETC &
	$0.0034$&$0.0025$&$-0.0012$&$49.4645$ \\
\bf July)	& BTC/ETH & 
	$0.0292$&$0.0198$&$0.0148$&$30.6927$ \\
	& BTC/LTC & 
	$0.0049$&$0.0036$&$-0.0004$&$46.1951$ \\
\midrule
\rowcolor{Gray}
\textbf{Kwon} 
	& BTC/BCH (ours) &
	$0.0089$&$0.0055$&$-0.0009$&$41.0030$ \\
\rowcolor{Gray}
\textbf{Range}
	& BTC/BCH \cite{Kwon:2019} &
	$ 0.0268$&$ 0.0196$&$-0.0155$&$31.4217$ \\
\bottomrule
\end{tabular}
}
\caption{\emph{Fit between equilibrium and actual allocation.
Root mean square error, mean average error, mean error (all in allocation units; lower is better) and peak signal-to-noise ratio (in dB; higher is better) for hourly data. Partial data for the given time period is indicated with an asterisk. The \emph{Kwon Range} compares our results to those of Kwon et al.~\cite{Kwon:2019} for an overlapping period ranging through most of 2018.
}}
\label{tab:gof-indicators}
\vspace{-3ex}
\end{table}

Agreement between $\boldsymbol{w}$ and $\boldsymbol{w}_{\texttt{eq}}$ in the ETH / ETC plot is not quite as strong as in BTC / BCH. There are two major deviations between actual and equilibrium allocations. First, there are several prominent spikes evident in the equilibrium. These all originate from excessively large transaction fees in ETC, which we believe appeared too suddenly for miners to respond by reallocating hash power. The most prominent occurs around the time of a known attack on ETC~\cite{etc51pct:2019}. After removing the top 0.1\% largest fees, the spikes disappear. Second there is a subtle bias toward ETH in the equilibrium for the dates prior to May 29, 2018, the date of a hard fork on the ETC chain, after which the bias abruptly vanishes. The hard fork removed a {\em difficulty bomb}, a piece of code that intentionally increases the difficulty (and therefore block time), so as to encourage a hard fork. 
However, the difficulty bomb does not affect the calculation of nominal hash rate (several bombs active for ETH during this time period have no effect), and the bomb was not active before mid February 2018 even though the bias existed earlier than that. So we cannot identify a definitive reason for this early bias. Figure~\ref{fig:me-months-eth-etc} in Appendix~\ref{sec:suppl_figs} shows the trend in overall error between actual and equilibrium allocation for the ETH / ETC pair.

Kwon et al.~\cite{Kwon:2019} also considered convergence of security allocation (hash rate) to the equilibrium among blockchains utilizing the same PoW algorithm. However, their analysis was considerably more limited. Their dataset focussed exclusively on the SHA256 PoW algorithm and was limited to dates prior to 2019. Furthermore, it failed to account for some protocol nuances such as transaction fees and bias in BTC's nominal hash rate. As a result, their findings conveyed much looser adherence to the equilibrium. Table~\ref{tab:gof-indicators} (gray) shows the error for BTC / BCH from January 1 to December 15, 2018 comparing the results of Kwon et al. with ours (the only dates our data overlapped with theirs). Note that we dropped dates from our data ranging from June 6 through August 12 because it was missing from their dataset. For root mean squared error, their results incur three times the error of ours. Moreover, mean error for their data reveals strong bias (most likely due to missing transaction fees).

\subsubsection{Multiple PoW Algorithms}
Miner adherence to the equilibrium is remarkably reliable between blockchains that share the same PoW algorithm. More remarkable still is that the equilibrium continues to hold between blockchains that do not share PoW algorithms. 

Similar to Figure~\ref{fig:single_pow_eq}, Figure~\ref{fig:multi_pow_eq} plots actual security allocation $\boldsymbol{w}$, in blue, and equilibrium allocation $\boldsymbol{w}_{\texttt{eq}}$, in red, this time for pairs of blockchains BTC / ETH and BTC / LTC. Because these plots pair blockchains that do not share a PoW algorithm, arbitrage must be achieved by trading hash rate through a secondary market such as NiceHash. The plots show generally good agreement, in terms of both magnitude and correlation between curves, but the equilibrium allocation in both plots does reveal significant bias during the month of January and again after May, 2020. Table~\ref{tab:gof-indicators} shows that the equilibrium for BTC / LTC deviates from the actual allocation with overall error of the same order as was observed for single PoW pairs. In contrast, deviation between the equilibrium and actual allocation of BTC / ETH shows error roughly 10 times greater than that of single PoW pairs. Bias is similarly elevated. Nevertheless, both root mean square error and mean square error remain below 3\%. 

Overall, the results suggest that \emph{the Blockchain Security Market seeks the point of no arbitrage even if it can only be accessed through secondary hash rate markets}. We hypothesize that it is inefficiency in the the hash rate market itself that introduces the higher error in the agreement between equilibrium and actual allocations.

%% file: econometrics.tex
% !TEX root = main.tex

\section{Causal Analysis}
\label{sec:econ}

Section~\ref{sec:eval} depicts a very strong historical correlation between security allocation and the allocation equilibrium predicted by Theorem~\ref{thm:equil}. The demonstration is empirical, but graphical nonetheless. In this section, we dig deeper into the relationship between actual and equilibrium allocations among chains sharing a PoW algorithm. Specifically, we ask, \emph{to what extent does change in actual allocation invoke change in the equilibrium, and vice versa}? 
Actual security allocation $\boldsymbol{w}$ is a function of the hash rate that miners devote to each chain, while the equilibrium allocation $\boldsymbol{w}_{\texttt{eq}}$ is a function of the fiat exchange price of the coins native to those blockchains. Thus, the question asks how hash rate and coin price mutually influence each other. 

By evaluating {\em Granger causality}~\cite{Granger:1969} between these quantities, we find strong evidence that coin price influences hash rate allocation, which implies that miner security allocation follows the equilibrium. However, the opposite is not typically true: hash rate allocation rarely exhibits a causal effect on coin price. This is not to say that increased hash rate cannot move coin price, only that we find scant evidence for it on a systematic, hourly basis. One possible reason for a lack of observed (Granger) causal effect may be because small changes in hash rate away from the equilibrium will be quickly offset by other miners exploiting the arbitrage opportunity it creates.

These findings have profound implications for blockchain security and governance. First, they imply that blockchains with fixed coin issuance and low coin value are destined to suffer from commensurately low security so long as their coin's price is suppressed. Second, dramatic changes in coin price, which are commonly observed in the cryptocurrency realm, can cause equally sudden changes in security. Third, we find little evidence that security improvements (reductions) are rewarded (punished) by the market. This does not rule out the possibility that it happens occasionally, but the signal is typically very weak. Indeed, individual market participants (i.e., miners) may attempt to improve security by increasing  security allocation to a given chain. But their efforts tend to be too marginal or their effort is offset by reverse actions of other market participants who exploit arbitrage and bring security back toward equilibrium.

Granger causality is a method used to establish causality between two time series with the simple rationale that a later event cannot give rise to an earlier one.
This notion of causality is weaker than the ``gold standard'' obtained from controlled experiments, which are very difficult to conduct in real markets. 
Granger causality assumes that there exists no unobserved third variable influencing events in both series with different latency.
With this caveat in mind, we proceed by estimating pairs of regression equations, each with a time-lagged version of the other as a predictor.

Regression on a time series amounts to extracting a stochastic process from temporal data. As a byproduct of the temporal nature of the data, standard regression techniques can often lead to dependent residuals, which imply the process is non-stationary, compromising the validity of statistical inference~\cite{Sargan:1983}. The problem manifests with the existence of unit (i.e., trivial) roots in the characteristic regression equation.
The standard solution is to differentiate the dependent variables in the equation several times until the unit-root vanishes.
Table~\ref{tab:adf-test} (see Appendix~\ref{sec:suppl_figs}) shows that this happens after calculating first differences for all our series of interest.
This implies that, while raw security allocations and equilibrium points are not stationary, hourly changes in these variables are.
Therefore the analyses in this section refer to first differences of series calculated from empirical data.
This transformation does not affect the logic behind Granger causality.

To determine if change in coin exchange price (labeled \emph{price change}), as the main component of the equilibrium, Granger-causes actual security allocation rebalancing (labeled \emph{security rebalancing}), we fit the following two specifications,
\begin{align}
	\Delta w_t + \varepsilon_t&= a + b_1 \cdot \Delta w_{t-1}  \label{eq:gc1} \\
	\Delta w_t + \varepsilon_t&= a + b_1 \cdot \Delta w_{t-1} + b_2 \cdot \Delta {w_\text{eq}}_{t-1}, \label{eq:gc2}  
\end{align}
and test if the additional term related to coefficient $b_2$ in Specification~\ref{eq:gc2} improves  explanatory power over Specification~\ref{eq:gc1}.
Likewise, we check if security rebalancing Granger-causes change in equilibrium allocation (a proxy for price change) by fitting
\begin{align}
	\Delta  {w_\text{eq}}_{t}+ \varepsilon_t&= a + b_1 \cdot \Delta {w_\text{eq}}_{t-1}  \label{eq:gc3} \\
	\Delta  {w_\text{eq}}_{t} + \varepsilon_t&= a + b_1 \cdot \Delta {w_\text{eq}}_{t-1} + b_2 \cdot \Delta w_{t-1}, \label{eq:gc4}  
\end{align}
and performing the same test.
All models are fit by minimizing the squares of the residuals $\varepsilon_t$. 

\begin{figure}

\providecommand{\GCabsent}{{\color{black}\large $\bullet$}}
\providecommand{\GCmarginal}{{\color{gray}\large $\bullet$}}
\providecommand{\GCweak}{{\color{yellow}\large $\bullet$}}
\providecommand{\GCmoderate}{{\color{green}\large $\bullet$}}
\providecommand{\GCstrong}{{\color{green!75!black}\large $\bullet$}}

\textbf{BTC / BCH}\\[-2ex]

{\centering
Price change causes security rebalancing

\begin{tikzpicture}[x=8pt]
	\input{econometrics/fig/gc-forward-btc-bch}
	\foreach \x/\q in {-.5/2018,11.5/2019,23.5/2020}
		\draw (\x,-8pt) --++(0,16pt);
\end{tikzpicture}

Security rebalancing causes price change

\begin{tikzpicture}[x=8pt]
	\input{econometrics/fig/gc-reverse-btc-bch}
	\foreach \x/\q in {-.5/2018,11.5/2019,23.5/2020}
		\draw (\x,-20pt) node [above right] {\small\q} --++(0,28pt);
\end{tikzpicture}
}

\hrule
\vspace{1em}

% ETH /ETC

\textbf{ETH / ETC}\\[-2ex]

{\centering
Price change causes security rebalancing

\begin{tikzpicture}[x=8pt]
	\input{econometrics/fig/gc-forward-eth-etc}
	\foreach \x/\q in {-.5/2018,11.5/2019,23.5/2020}
		\draw (\x,-8pt) --++(0,16pt);
\end{tikzpicture}

Security rebalancing causes price change

\begin{tikzpicture}[x=8pt]
	\input{econometrics/fig/gc-reverse-eth-etc}
	\foreach \x/\q in {-.5/2018,11.5/2019,23.5/2020}
		\draw (\x,-20pt) node [above right] {\small\q} --++(0,28pt);
\end{tikzpicture}}

% CAPTION
\caption{\emph{Monthly results for the Granger causality analysis. The strength of the evidence for a causal link is color-coded by the statistical significance of the $F$-test as follows:
\GCabsent~\emph{absent} ($p>0.1$),
\GCmarginal~\emph{marginal} ($p\leq 0.1$),
\GCweak~\emph{weak}  ($p\leq 0.05$),
\GCmoderate~\emph{moderate} ($p\leq 0.01$), and
\GCstrong~\emph{strong} ($p \leq 0.001$).
Input data: hourly first differences.
}}
\label{fig:causality}
\end{figure}
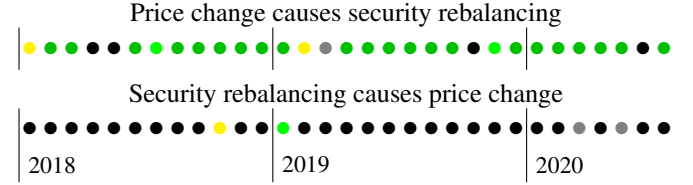

Figure~\ref{fig:causality} shows the strength of Granger-causal link for blockchain pairs BTC / BCH (top facet) and ETH / ETC (bottom facet) from price change to security rebalancing (top row) and security rebalancing to price change (bottom row). In particular, the figure reports the probability of a type I error in choosing Specification~\ref{eq:gc2} (top row) over~\ref{eq:gc1} and choosing Specification~\ref{eq:gc4} over~\ref{eq:gc3} (bottom row). Dark green dots indicate very low $p$-value, or a strong Granger-causal link, while black dots indicate a very high $p$-value, or no Granger-causal link.

On a systematic, hourly basis, we find overwhelming evidence that security rebalancing follows price change. But only in rare circumstances will the market look to hash rate (security rebalancing) to readjust price. This lends strong support to the conclusion that change in coin price (and thus expected reward) Granger-causes security rebalancing, but typically not the opposite. One reason for the rarity of a Granger-causal link in the opposite direction might be that a change in security allocation away from the equilibrium is quickly offset by other miners shifting their hash rate in the opposite direction.
Nevertheless, the conditions under which security rebalancing \emph{does} influence price change are noteworthy. We can see from Figure~\ref{fig:causality} (top) that the only month that saw a strong link for the BTC / BCH pair was during the last subsidy halving for BTC, when the profitability of BTC coins was suddenly cut in half. For the ETH / ETC pair, security rebalancing was moderately predictive of price change only during a month when ETC experienced a series of 51\% attacks~\cite{etc51pct:2019}.

We leave deeper Granger-causal analysis for future work including more variables (if observable) and a broader class of specifications.

%% file: econometrics/fig/gc-forward-btc-bch.tex
\draw (0,0) node { \GCmoderate }++(1,0)node { \GCmoderate }++(1,0)node { \GCstrong }++(1,0)node { \GCstrong }++(1,0)node { \GCstrong }++(1,0)node { \GCmoderate }++(1,0)node { \GCabsent }++(1,0)node { \GCstrong }++(1,0)node { \GCweak }++(1,0)node { \GCstrong }++(1,0)node { \GCstrong }++(1,0)node { \GCstrong }++(1,0)node { \GCabsent }++(1,0)node { \GCabsent }++(1,0)node { \GCmoderate }++(1,0)node { \GCstrong }++(1,0)node { \GCstrong }++(1,0)node { \GCstrong }++(1,0)node { \GCmoderate }++(1,0)node { \GCstrong }++(1,0)node { \GCweak }++(1,0)node { \GCmoderate }++(1,0)node { \GCweak }++(1,0)node { \GCmoderate }++(1,0)node { \GCmoderate }++(1,0)node { \GCstrong }++(1,0)node { \GCabsent }++(1,0)node { \GCstrong }++(1,0)node { \GCabsent }++(1,0)node { \GCabsent }++(1,0)node { \GCweak } ;

%% file: econometrics/fig/gc-reverse-btc-bch.tex
\draw (0,0) node { \GCabsent }++(1,0)node { \GCabsent }++(1,0)node { \GCabsent }++(1,0)node { \GCabsent }++(1,0)node { \GCabsent }++(1,0)node { \GCabsent }++(1,0)node { \GCweak }++(1,0)node { \GCabsent }++(1,0)node { \GCabsent }++(1,0)node { \GCabsent }++(1,0)node { \GCabsent }++(1,0)node { \GCabsent }++(1,0)node { \GCabsent }++(1,0)node { \GCabsent }++(1,0)node { \GCabsent }++(1,0)node { \GCmarginal }++(1,0)node { \GCabsent }++(1,0)node { \GCabsent }++(1,0)node { \GCabsent }++(1,0)node { \GCabsent }++(1,0)node { \GCabsent }++(1,0)node { \GCabsent }++(1,0)node { \GCabsent }++(1,0)node { \GCabsent }++(1,0)node { \GCabsent }++(1,0)node { \GCabsent }++(1,0)node { \GCabsent }++(1,0)node { \GCabsent }++(1,0)node { \GCstrong }++(1,0)node { \GCabsent }++(1,0)node { \GCweak } ;

%% file: econometrics/fig/gc-forward-eth-etc.tex
\draw (0,0) node { \GCweak }++(1,0)node { \GCstrong }++(1,0)node { \GCstrong }++(1,0)node { \GCabsent }++(1,0)node { \GCabsent }++(1,0)node { \GCstrong }++(1,0)node { \GCmoderate }++(1,0)node { \GCstrong }++(1,0)node { \GCstrong }++(1,0)node { \GCstrong }++(1,0)node { \GCstrong }++(1,0)node { \GCstrong }++(1,0)node { \GCstrong }++(1,0)node { \GCweak }++(1,0)node { \GCmarginal }++(1,0)node { \GCstrong }++(1,0)node { \GCstrong }++(1,0)node { \GCstrong }++(1,0)node { \GCstrong }++(1,0)node { \GCstrong }++(1,0)node { \GCstrong }++(1,0)node { \GCabsent }++(1,0)node { \GCmoderate }++(1,0)node { \GCstrong }++(1,0)node { \GCstrong }++(1,0)node { \GCstrong }++(1,0)node { \GCstrong }++(1,0)node { \GCstrong }++(1,0)node { \GCstrong }++(1,0)node { \GCabsent }++(1,0)node { \GCstrong } ;

%% file: econometrics/fig/gc-reverse-eth-etc.tex
\draw (0,0) node { \GCabsent }++(1,0)node { \GCabsent }++(1,0)node { \GCabsent }++(1,0)node { \GCabsent }++(1,0)node { \GCabsent }++(1,0)node { \GCabsent }++(1,0)node { \GCabsent }++(1,0)node { \GCabsent }++(1,0)node { \GCabsent }++(1,0)node { \GCweak }++(1,0)node { \GCabsent }++(1,0)node { \GCabsent }++(1,0)node { \GCmoderate }++(1,0)node { \GCabsent }++(1,0)node { \GCabsent }++(1,0)node { \GCabsent }++(1,0)node { \GCabsent }++(1,0)node { \GCabsent }++(1,0)node { \GCabsent }++(1,0)node { \GCabsent }++(1,0)node { \GCabsent }++(1,0)node { \GCabsent }++(1,0)node { \GCabsent }++(1,0)node { \GCabsent }++(1,0)node { \GCabsent }++(1,0)node { \GCabsent }++(1,0)node { \GCmarginal }++(1,0)node { \GCabsent }++(1,0)node { \GCmarginal }++(1,0)node { \GCabsent }++(1,0)node { \GCabsent } ;

%% file: oracle.tex
% !TEX root = main.tex

\section{Trustless Price-Ratio Oracle}
\label{sec:oracle}

Price oracles are a fundamental tool for many popular smart contract applications~\cite{gnosis,Augur,Numerai}, particularly in the nascent space of decentralized finance (DeFi)~\cite{dydx,Compound,Dharma}. Oracles typically pull data from trusted, centralized sources~\cite{Makerdao,Provable,coinbase}, decentralized exchanges~\cite{Uniswap,bancor}, or from crowds~\cite{whgeorge,DutchX}. Centralized sources require trust in a corruptible third-party, while crowd sourcing and decentralized exchanges are subject to manipulation. For example lending platform bZx recently lost the equivalent of nearly \$1M USD due to exchange price manipulation~\cite{bZxHack}. 

In this section, we describe a smart contract $\texttt{Oracle}$ that leverages the allocation equilibrium described by Theorem~\ref{thm:equil} to provide an estimate of the fiat \emph{price ratio} of the cryptocurrencies $A$ and $B$ from information contained in block headers only. It can either be used alone or aggregated with existing solutions to increase robustness. An example of a futures contract leveraging $\texttt{Oracle}$ appears in Appendix~\ref{sec:fut_example}. 

In estimating the price ratio of two coins sharing the same PoW algorithm, $\texttt{Oracle}$ can be no more easily manipulated than the PoW that secures each of the chains. For coins that use different PoW algorithms, Section~\ref{sec:oracle_different} describes a contract called $\texttt{Spot}$, which is necessary for determining the ratio of spot hash prices. Although $\texttt{Spot}$ is susceptible to manipulation, we explain below that its extent is quantifiable and limited. 

Smart contract $\texttt{Oracle}$ runs on chain $A$, returning an estimate of the price ratio $P_B / P_A$ when the two chains are each at a given block height. It does so by implementing a light client for blockchain $B$.  Two public methods are exposed: $\texttt{Update}(h_B)$ and $\texttt{Query}(b_A, b_B, \sigma_{\Delta})$ (see Algorithms~\ref{alg:oracle_update} and~\ref{alg:oracle_query} in Appendix~\ref{sec:algorithms}). Method $\texttt{Update}(h_B)$ allows any user to update the chain of headers with a new header $h_B$ having the following properties: 
\1 the previous block hash of $h_B$ points to the block hash of the previous header; and 
\2 the PoW associated with the hash of $h_B$ meets the difficulty implied by earlier headers and chain $B$'s protocol. If either of the conditions are not met, then $\texttt{Update}$ returns an error. 

Method $\texttt{Query}(b_A, b_B, \sigma_\Delta)$ returns an estimate of the price ratio $P_B / P_A$ at the time when chain $A$ was at block height $b_A$, chain $B$ was at height $b_B$, and the ratio of spot hash prices is equal to $\sigma_\Delta$, i.e., $\sigma_\Delta = \frac{\sigma_B}{\sigma_A}$. If either \1 the header at block height $b_B$ is unknown to $\texttt{Oracle}$ or \2 the block on chain $A$ at height $b_A$ has not yet been mined, then an error is thrown. We assume  any party interested in querying the oracle will be incentivized to run $\texttt{Update}(h_B)$ for all new 
headers $h_B$.

The initial state of contract $\texttt{Oracle}$ is comprised of list $\texttt{Headers}_B = [h_B^*]$ where $h_B^*$ is the header for the genesis block on chain $B$. The latest list of headers,  $\texttt{Headers}_A$, is native to blockchain $A$ and is therefore assumed to be accessible from within $\texttt{Oracle}$. Furthermore, let $h_X[g]$, $h_X[D]$, and $h_X[P]$ denote the target, difficulty, and hash of previous block, respectively, specified in header $h_X$ for $X \in \{A, B\}$. Finally, define $\texttt{Headers}_X[\minus 1]$ to be the last item added to list $\texttt{Headers}_X$ and let $\mathcal{H}_X(h_X)$ denote the hash of header $h_X$.

Using difficulties $D_A$ and $D_B$, extracted from the headers on chains $A$ and $B$, and spot hash price ratio $\sigma_\Delta$, $\texttt{Oracle}$ estimates $P_B / P_A$ by equating $w_A$ from Eq.~\ref{eq:infer_w} and $w_A$ from Theorem~\ref{thm:equil}: \vspace{-1ex}
\[
w_A \approx \frac{\hat{s}_A}{\hat{s}_A + \hat{s}_B},
\]
where $w_A$ denotes the portion of allocation among chains $A$ and $B$ devoted to chain $A$ and $\hat{s}_X = \sigma_X \frac{D_X}{T_X}$. It follows that,\vspace{-1.0em}
\begin{equation}
\label{eq:oracle}
\renewcommand*{\arraystretch}{2.0}
\begin{array}{l}
\frac{\hat{s}_A}{\hat{s}_A + \hat{s}_B} \approx \frac{T_B R}{T_B R - T_A R + T_A} \Rightarrow \\
\frac{P_B}{P_A} \approx \frac{k_A}{k_B T_A} \left( \frac{T_B (\hat{s}_A + \hat{s}_B) }{\hat{s}_A} - T_B + T_A \right) - \frac{k_A}{k_B} \\
\frac{P_B}{P_A} \approx \frac{k_A T_B}{k_B T_A} \frac{\hat{s}_B}{\hat{s}_A} =  \frac{k_A}{k_B} \frac{D_B}{D_A} \frac{\sigma_B}{\sigma_A} = \sigma_\Delta \frac{k_A}{k_B} \frac{D_B}{D_A}.
\end{array}
\end{equation}

\subsection{Oracle for Common PoW Algorithms}

When blockchains $A$ and $B$ use the same PoW algorithm, $\sigma_\Delta = 1$, and Eq.~\ref{eq:oracle} simplifies to $\frac{P_B}{P_A} \approx \frac{k_A}{k_B} \frac{D_B}{D_A}$. In this case, all information required by $\texttt{Oracle}$ is either provided to the contract by way of the $\texttt{Update}$ method or is accessible natively on chain $A$.

Figure~\ref{fig:single_pow_eq} and the MAE from Table~\ref{tab:gof-indicators} demonstrate that the equilibrium agrees strongly with the actual allocation when chains $A$ and $B$ use the same PoW algorithm. Thus, we can verify price-ratio predictions would have been accurate within less than 1\% error. 

\subsection{Oracle for Different PoW Algorithms}
\label{sec:oracle_different}

Smart contract $\texttt{Oracle}$ cannot be applied directly when chains $A$ and $B$ implement different PoW algorithms because spot hash prices do not cancel from Eq.~\ref{eq:security}. As a result, we must use Eq.~\ref{eq:infer_w} to approximate relative security. 
In this section, we show that it is possible to approximate the ratio of coin prices $\frac{P_B}{P_A}$ with knowledge only of the ratio $\sigma_\Delta = \frac{\sigma_A}{\sigma_B}$, which we presently show how to calculate. To that end, we describe a new smart contract, $\texttt{Spot}$, that serves to estimate the spot hash price ratio, which can be used with $\texttt{Oracle}$ when PoW algorithms differ between chains. 

$\texttt{Spot}$ implements a variation of the scheme of Luu et al.~\cite{Luu:2016}. Every \emph{epoch}, a reward is offered for solving a mining puzzle equivalent to PoW algorithm $\texttt{ALG}_B$, only with a customized target. Each epoch has multiple \emph{rounds}, and the puzzle target changes every round. Suppose that epoch $i$ generated $r_i$ rounds, and let $g_{r_i}$ denote the final target in epoch $i$, which was achieved in round $r_i$. In round 1 of epoch $i+1$, the target is set to $\alpha g_{r_i}$, where $\alpha > 0$ is also a tunable parameter, and miners are invited to solve the puzzle in return for reward $\mathcal{R}$, paid in units of coin $A$. If after $N$ blocks on chain $A$ ($N$ being similarly tunable) no solution to the puzzle has been submitted, then the target is raised by \emph{target step} $g_{\Delta}$ and round 2 commences. The step is defined as $g_{\Delta} = g_{r_i} / j$, where $ j > 0$ is a tunable parameter. The process continues until the final round when a valid solution is submitted to the contract.

Given previous target $g_{r_i}$ and the final round count $r_{i+1}$, along with fixed parameters $j$, $\alpha$, and $N$, we can estimate the spot price ratio $\sigma_A / \sigma_B$ for epoch $i+1$. Ozisik et al.~\cite{Ozisik:july2017} showed that the expected number of hashes $H$ performed in mining a block (or solving a mining puzzle) with target $g$ is given by $H = \mathcal{S} / g$, where $\mathcal{S}$ is the size of the hash space. It follows that, in expectation, $H_{i+1} = \mathcal{S} / (\alpha g_i + r_{i+1} g_\Delta)$ hashes were performed during epoch $i+1$ in computing the hash inversion puzzle paying reward $\mathcal{R}$. This implies that each hash using the PoW algorithm of chain $B$ is worth $\mathcal{R} (\alpha g_i + r_{i+1} g_\Delta) / \mathcal{S}$ units of chain $A$ coin. Now suppose that, at the same time, the target on chain $A$ is $g_A$. The reward per unit hash for mining a block on chain $A$ is given by $k_A  g_A / \mathcal{S}$, where $k_A$ is the number of coins awarded for mining a block on chain $A$. An economically rational miner capable of producing hashes from algorithm $\texttt{ALG}_B$ will therefore place the following value on the spot price ratio.\vspace{-0em}
\begin{equation}
\sigma_\Delta = 
\frac{k_A g_A / \mathcal{S}}{\mathcal{R} (\alpha g_i + r_{i+1} g_\Delta) / \mathcal{S}} = \frac{k_A g_A}{\mathcal{R} (\alpha g_i + r_{i+1} g_\Delta)}.\label{eq:inferred_splot_price}
\end{equation}

There are 2 public methods on contract $\texttt{Spot}$: $\texttt{Solve}(a, \texttt{n})$ and $\texttt{Query}(f)$ and one private method $\texttt{Update}()$, which runs automatically every time a block is produced on chain $A$ (see Algorithms~\ref{alg:spot_solve}, \ref{alg:spot_query},  and~\ref{alg:spot_update} in Appendix~\ref{sec:algorithms}). Contract state comprises parameters $j$, $\alpha$, $N$, and the following variables. Reward $\mathcal{R} = 0$; fee balance $\mathcal{F} = 0$; round counter $r = 1$; round target $g_r = \mathcal{S}$; round target change $\Delta g = \mathcal{S} / j$; and block counter $b_A = \texttt{length}(\texttt{Headers}_A)$. The contract additionally stores prior target $g' = \mathcal{S}$ and target change $\Delta g' = \mathcal{S}/j$. We assume that the complete list of headers, $\texttt{Headers}_A$, is natively accessible. For each header $h_A \in \texttt{Headers}_A$, let $h_A[g]$ denote the target.

Method $\texttt{Solve}(a, \texttt{n})$ accepts payout address $a$ and solution nonce $\texttt{n}$. If the solution is valid, then $\texttt{Spot}$ updates contract state to reflect the target at which the puzzle was solved and deposits $\mathcal{R}$ coins into account $a$. If the solution is not valid, then no action is taken. Method $\texttt{Query}(f)$ accepts only quantity $f$ of coin $A$ as fee, and returns the latest calculation of spot ratio according to Eq.~\ref{eq:inferred_splot_price}. The funds comprising $\mathcal{R}$ are derived from fees paid by participants who use the service by calling $\texttt{Query}$. For simplicity, we assume 
$\mathcal{R}$ remains fixed, but it could 
be set to a fixed fraction of the remaining fees collected.

\para{Manipulation.} Notice that, according to Eq.~\ref{eq:inferred_splot_price}, the spot price ratio $\Delta \sigma$ can be manipulated by a miner with exogenous economic motivations who devotes \emph{more} hash rate to solving the puzzle than would ordinarily be profitable given the value of coin $A$ relative to coin $B$. This results in a solution after \emph{fewer} rounds than expected, which leads to an artificial increase in the spot price of hashes for $\texttt{ALG}_A$ over $\texttt{ALG}_B$. Yet, any hash rate diverted to solving the puzzle will come at the expense of mining blocks on chain $B$. Thus, the threat of manipulation can be indirectly quantified. Moreover, it is much more difficult to artificially \emph{decrease} $\Delta \sigma$ because it would require all miners to abstain from solving the puzzle even when it is more profitable than mining chain $B$. Accordingly, a variation of $\texttt{Spot}$ could be implemented either on both chains $A$ and $B$ or both on chain $A$ alone with one reward denominated in coin $A$ and the other in coin $B$. The true value of $\Delta \sigma$ could be taken as the minimum of the two reported values, making it 
more difficult to manipulate.

%% file: related.tex
% !TEX root = main.tex

\section{Related Work}
\label{sec:related}

In the context of a single blockchain, Prat and Walter~\cite{Prat:2018} model the impacts of mining difficulty and coin exchange rate on profitability. 
Ma et al.~\cite{June:2019} show that there exists a Nash equilibrium for the computing power allocated by miners given a fixed mining difficulty. Kristoufek~\cite{Kristoufek:2020} derives an equilibrium between miner hash rate production and PoW energy costs in Bitcoin mining. 
Huberman et al.~\cite{Huberman:2019} devise an economic model of the Bitcoin payment system that captures the tension between users who compete for transaction processing capacity provided by miners. 
Biais et al.~\cite{Biais:2019} identify Markov-perfect equilibria in miner consensus; their analysis is primarily theoretical with only anecdotal supporting evidence.

Huang et al.~\cite{Huang:2018} describe short-term investing and mining strategies for cryptocurrencies relative to base currencies Litecoin and Bitcoin. Nguyen et al.~\cite{Nguyen:2019} show that new cryptocurrencies have a small but significant negative impact on the price of Bitcoin. Both stop short of identifying hash rate allocation equilibria.
Gandal et al.~\cite{Gandal:2018} analyzes price manipulation on the Mt. Gox exchange, concluding that it was carried out by a small group. Today there exist many exchanges, centralized and decentralized, which makes such manipulation %much 
more difficult.

Meshkov et al.~\cite{Meshkov:2017} 
analyze \emph{coin-hopping}, where miners move among blockchains using the same PoW according to which is most profitable;  
see also~\cite{Kiraly:2018}. 
Several works 
determine the optimal hash rate allocation between blockchains for \emph{individual} miners or mining pools; e.g.~\cite{Bissias:2018,Cong:2018,Chatzigiannis:2019}.

Spiegelman et al.~\cite{Spiegelman:2018} apply the theory of Potential Games~\cite{Monderer:1996} to the problem of miner hash rate allocation across multiple blockchains. They prove that multiple stable equilibria can exist, and that they can be achieved without the use of a sophisticated utility function. However, they provide no means to explicitly identify equilibria, nor is it clear from their work how a single equilibrium is achieved among the multiple possibilities. 
Altman et al.~\cite{Altman:2018} reach similar conclusions using a different model of hash rate allocation across cryptocurrencies and mining pools.

Kwon et al.~\cite{Kwon:2019} show that there exist multiple Nash equilibria for miners who allocate their hash rate among two blockchains sharing the same PoW. One of their equilibria coincides with $\boldsymbol{w}_{\texttt{eq}}$, the equilibrium we study, which they demonstrate is observed in practice. However they fail to provide adequate justification for its formation or uniqueness among other identified equilibria. The utility function used in their analysis is quite complex and incorporates knowledge of the hash power of miners, which is not publicly available. 

Han et al.~\cite{Han:2019} investigate doublespending on blockchains with relatively low hash rate instigated by either miners from a higher hash rate chain or attackers who purchase hash rate from a marketplace such as NiceHash~\cite{nicehash}. 
Sapirshtein et al.~\cite{Sapirshtein:2015} and Gervais et al.~\cite{Gervais:2016} apply MDPs to blockchains to analyze selfish mining~\cite{Eyal:2014} and double spend attacks. 

There is much existing work in the finance literature related to \emph{no arbitrage} (NA) conditions and the \emph{law of one price} (LOOP) in the presence of short sale restrictions.  
Discrete-time models include: LeRoy et al.~\cite{LeRoy:2014}, Chichilnisky~\cite{Chichilnisky:1995}, He et al.~\cite{He:1991}, and Oleaga~\cite{Oleaga:2012}. And continuous-time formulations include: Napp~\cite{Napp:2003}, Pulido~\cite{Pulido:2014}, Coculescu et al.~\cite{Coculescu:2017}, and Jarrow et al.~\cite{Jarrow:2015}. Continuous-time models are much richer than what is necessary for our work, thus we use a discrete-time model in this document. 
Kroeger and Sarkar~\cite{Kroeger:2017} show that the LOOP does \emph{not} hold in the Bitcoin / fiat exchange market, indicating frictions in some of these marketplaces. 
Yaish and Zohar~\cite{Yaish:2020} use the NA principle to price ASIC mining hardware.

%% file: conclusion.tex
% !TEX root = main.tex

\section{Conclusion}

We have presented a novel theory of the fiat value of security allocated among PoW blockchains, which is supported with empirical evidence
and novel applications. Our principle finding is that, for any pair of cryptocurrencies, not necessarily sharing the same PoW algorithm, there exists a unique equilibrium allocation, based on market prices only, that is robust even to intentional manipulation of miner hash rate. We furthermore establish a strong Granger-causal link from market price change to change in security allocation, the opposite link is found to hold only under exceptional circumstances. We end with a trustless price ratio oracle that leverages the allocation equilibrium. The generality of our framework opens new doors for future work. In particular, our characterization of security in terms of opportunity cost can generalize to other consensus principles, such as PoS.

%% file: symbols.tex
% !TEX root = main.tex

\vspace{-.5em}
\section{List of Symbols and Supplemental Figures}
\label{sec:suppl_figs}

\noindent{\relsize{-1}
\begin{tabular*}{.99\columnwidth}{c l}
\toprule
{\bf Symbol } &~~ {\bf Description} \\
\cmidrule(lr){1-2} \cmidrule(lr){1-2}
$A, B$ &~~ Either an arbitrary blockchain (chain) or its  native coin  \\
$X$ &~~ Variable identifying a chain such that $X \in \{A, B\}$ \\
$\texttt{ALG}_X$ & ~~The PoW algorithm for chain $X$ \\
$M_X$ & ~~ The set of miners capable of performing PoW $W_X$ \\
$M$ & ~~ The union of miners $M_A$ and $M_B$ \\
$H_X$ & ~~ The number of $W_X$ hashes per second on chain $X$ \\
$T_X$ & ~~ Protocol targeted block inter-arrival time for	 chain $X$ \\
$t_X$ & ~~ Actual block inter-arrival time for chain $X$ \\
$\tau$ & ~~ Time since epoch \\
$\sigma_X$ & ~~ Fiat value of a single hash using $W_X$ \\
$S_X$ & ~~ Target security investment on chain $X$ \\
$s_X$ & ~~ Actual security investment on chain $X$ \\
$\boldsymbol{w}$ & ~~ Security allocation vector among chains $A$ and $B$ \\
$\boldsymbol{w}_i$ & ~~ Security allocation for $m_i$ among chains $A$ and $B$ \\
$\boldsymbol{u}_i$ & ~~ Share of reward on chains $A$ and $B$ for miner $m_i$ \\ 
$V_X$ & ~~ Fiat value of coinbase reward plus fees \\
$k_X$ & ~~ Number of coins in coinbase reward plus average fees \\
$P_X$ & ~~ Fiat value of a single coin from chain $X$ \\
$D_X$ & ~~ Difficulty, expected hashes required to mine a block\\
& ~~ on chain $X$ \\ 
$R$ & ~~ Relative reward for mining on chain $A$ \\
$N$ & ~~ Total number of miners in the Security Allocation Game \\
$m_i, m_{\minus i}$ & ~~ Miner $i$ and all other miners, respectively \\
$\boldsymbol{\pi}$ & ~~ Expected fiat payoff \\
$e, e_i$ & ~~ Initial fiat endowment in aggregate and for  miner $m_i$ \\
$\boldsymbol{c}, \boldsymbol{c}_i$ & ~~ Claim vector (of payoff) in aggregate and for $m_i$ \\
$\boldsymbol{p}$ & ~~ Portfolio pricing vector \\
$\Delta  \boldsymbol{w}$ & ~~ Allocation vector rebalancing \\
\bottomrule
\end{tabular*}
}

%% file: suppl_figs.tex
% !TEX root = main.tex

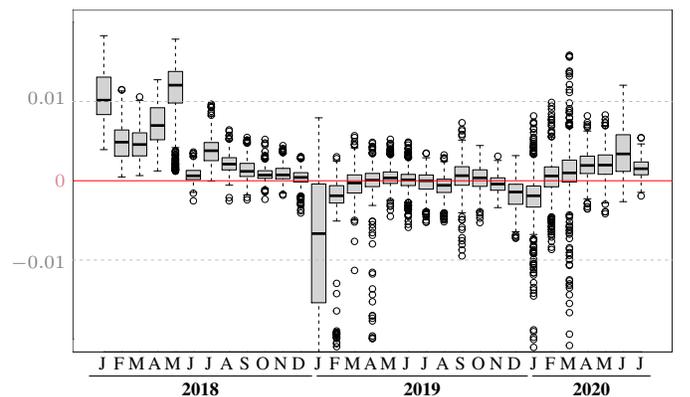
\begin{figure}[b!]
\begin{tikzpicture}
	\node [inner sep=0] (A) {\includegraphics[width=.9\linewidth]{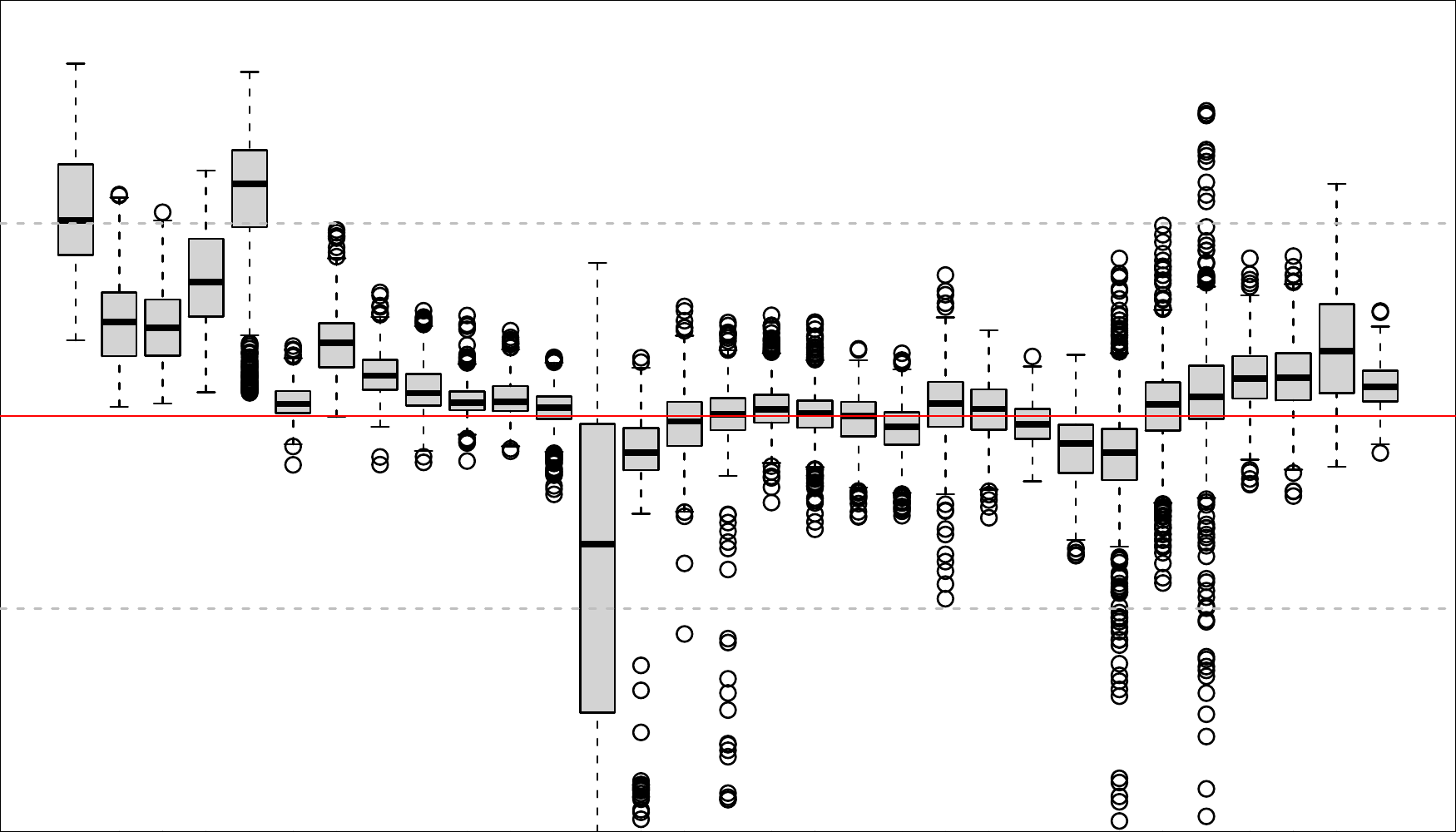}};
	\scriptsize
	\draw (A.west) node [left,red] {$0$};
	\draw [white] (A.west) -- node [pos=.48,left,black!50] {$-0.01$} (A.south west);
	\draw [white] (A.west) -- node [pos=.48,left,black!50] {$0.01$} (A.north west);

	\foreach \x in {0,12}
		\draw [x=2.39mm,anchor=base] (A.south west)++(\x,-2ex)++(1.6,0)
			node (J\x) {\,J} ++(1,0) node {F} ++(1,0) node {M} ++(1,0) node {A} ++(1,0) 
			node {M} ++(1,0) node {J} ++(1,0) node {J} ++(1,0) node {A} ++(1,0) 
			node {S} ++(1,0) node {O} ++(1,0) node {N} ++(1,0) node (D\x) {D\vphantom{J}} ++(1,0);
	\foreach \x in {24}
		\draw [x=2.39mm,anchor=base] (A.south west)++(\x,-2ex)++(1.6,0)
			node (J\x) {\,J} ++(1,0) node {F} ++(1,0) node {M} ++(1,0) node {A} ++(1,0) 
			node {M} ++(1,0) node {J} ++(1,0) node (JX) {J} ++(1,0); % node {A} ++(1,0) ;
			
	\draw (J0.south west)-- node [below] {\textbf{2018}} (D0.south east);
	\draw (J12.south)-- node [below] {\textbf{2019}} (D12.south east);
	\draw (J24.south)-- node [below] {\textbf{2020}} (JX.south east);
\end{tikzpicture}
\caption{\emph{Evolution of the ETH/ETC prediction error over time. 
Monthly distributions of hourly differences between equilibrium and actual hash rate allocation (in allocation units). 
Positive values mean that for the observed market prices, our theory predicts a higher allocation of hash power to ETH than estimated from block times.
The data ranges from January 1, 2018 until July 31, 2020.}}
\label{fig:me-months-eth-etc}
\end{figure}

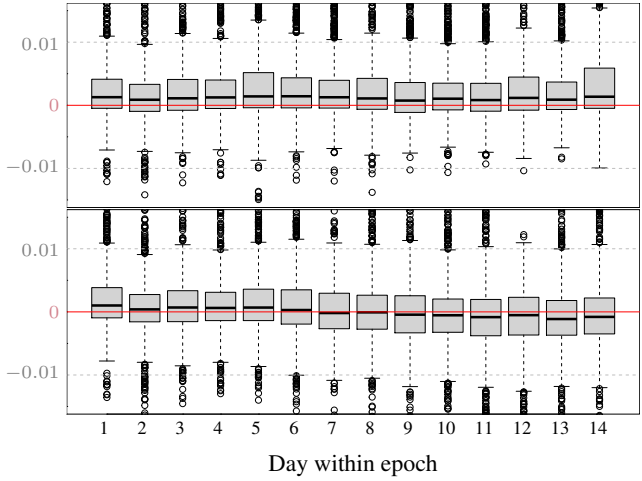
\begin{figure}[h!]
\begin{tikzpicture}
	\node [inner sep=0] (A) {\includegraphics[width=.86\linewidth]{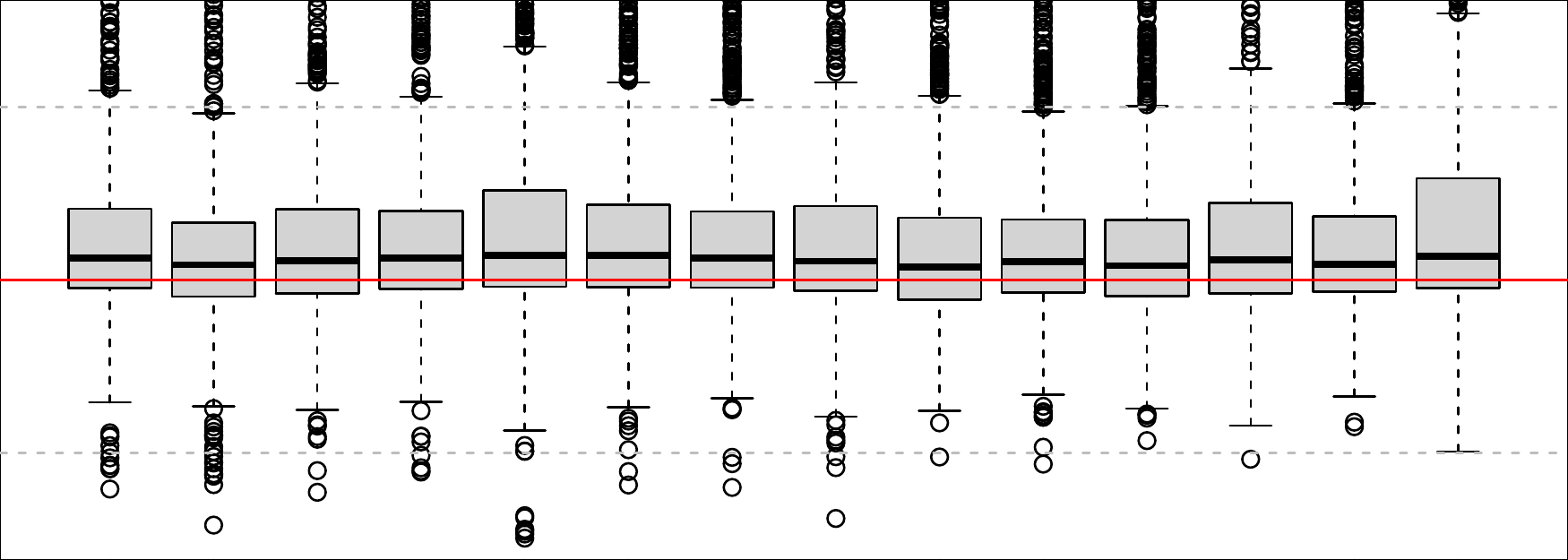}};
	\scriptsize
	\draw (A.west) node [left,red] {$0$};
	\draw [white] (A.west) -- node [pos=.6,left,black!50] {$-0.01$} (A.south west);
	\draw [white] (A.west) -- node [pos=.6,left,black!50] {$0.01$} (A.north west);
\end{tikzpicture}
\begin{tikzpicture}
	\node [inner sep=0] (A) {\includegraphics[width=.86\linewidth]{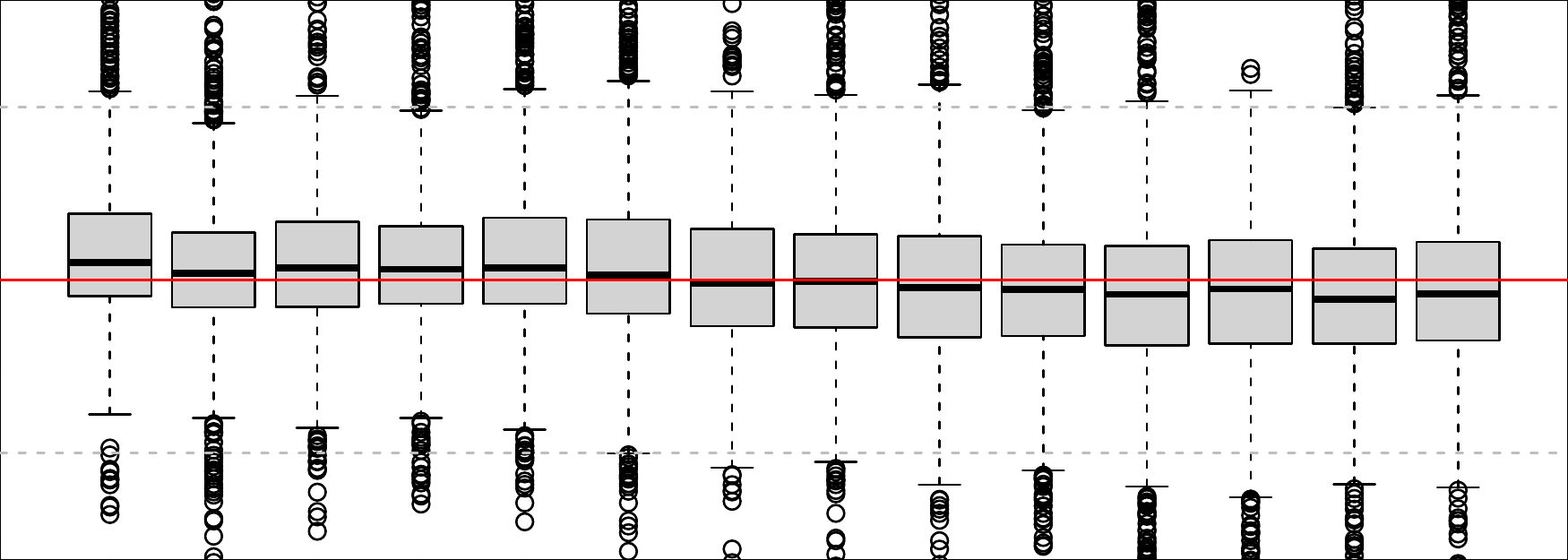}};
	\scriptsize
	\draw (A.west) node [left,red] {$0$};
	\draw [white] (A.west) -- node [pos=.6,left,black!50] {$-0.01$} (A.south west);
	\draw [white] (A.west) -- node [pos=.6,left,black!50] {$0.01$} (A.north west);
	\foreach \d in {1,2,...,14}
		\draw [x=5.05mm] (A.south west)++(\d,0) node [below] {\d};
	\small	
	\draw (A.south) node [below=3ex] {{Day within epoch}};
\end{tikzpicture}
\caption{\emph{Shift of prediction error within Bitcoin's SAA epoch. 
Distribution of hourly differences between equilibrium and actual hash rate allocation (in allocation units). 
Top: actual allocation estimated from the nominal difficulty for BTC and BCH.
Bottom: actual allocation estimated from a 96-hour moving window for BTC and nominal difficulty for BCH.}}

\label{fig:me-within-epoch}\vspace{0cm}
\end{figure}

\begin{table}[H]
\centering
\begin{tabular}{l rr rr}
\toprule
& \multicolumn{2}{c}{Raw}
& \multicolumn{2}{c}{1\textsuperscript{st} differences}\\
\cmidrule{2-3}
\cmidrule(l){4-5}
Time series 
& Statistic &\multicolumn{1}{c}{$p$} 
& Statistic &\multicolumn{1}{c}{$p$}  \\
\midrule
BTC/BCH actual
& \input{econometrics/fig/adf-btc-bch-allocation-level}
& \input{econometrics/fig/adf-btc-bch-allocation-diff}
\\
BTC/BCH equilibrium
& \input{econometrics/fig/adf-btc-bch-equilibrium-level}
& \input{econometrics/fig/adf-btc-bch-equilibrium-diff}
\\
ETH/ETC actual
& \input{econometrics/fig/adf-eth-etc-allocation-level}
& \input{econometrics/fig/adf-eth-etc-allocation-diff}
\\
ETH/ETC equilibrium
& \input{econometrics/fig/adf-eth-etc-equilibrium-level}
& \input{econometrics/fig/adf-eth-etc-equilibrium-diff}
\\
\bottomrule
\end{tabular}
\caption{\emph{Check of preconditions. 
Augmented Dickey--Fuller tests for unit-roots in the hourly time series used for Granger causality. 
Series with $p$-values rejecting the null hypothesis fulfill the conditions that make the asymptotic theory valid.}}
\label{tab:adf-test}
\end{table}

%% file: econometrics/fig/adf-btc-bch-allocation-level.tex
$-3.29$&$0.07$

%% file: econometrics/fig/adf-btc-bch-allocation-diff.tex
$-36.0$&$<0.01$

%% file: econometrics/fig/adf-btc-bch-equilibrium-level.tex
$-2.67$&$0.29$

%% file: econometrics/fig/adf-btc-bch-equilibrium-diff.tex
$-27.8$&$<0.01$

%% file: econometrics/fig/adf-eth-etc-allocation-level.tex
$-2.25$&$0.47$

%% file: econometrics/fig/adf-eth-etc-allocation-diff.tex
$-31.2$&$<0.01$

%% file: econometrics/fig/adf-eth-etc-equilibrium-level.tex
$-2.34$&$0.44$

%% file: econometrics/fig/adf-eth-etc-equilibrium-diff.tex
$-29.7$&$<0.01$

%% file: mdp_details.tex
% !TEX root = main.tex

\section{Markov Decision Process (MDP) Details}
\label{sec:mdp_details}

At a high level, our MDP is comprised of states corresponding to the current difficulty on each chain, actions correspond to the allocation of available hash rate between the two chains, and transitions occur every second. Details are given below.

\medskip

\noindent \textbf{States.} Each state is a tuple of the form $(D_A, D_B, \beta_A, \beta_B)$, where $D_A$ and $D_B$ are the difficulties on chains $A$ and $B$, respectively, given in terms of the expected number of hashes per block. Each chain is additionally given a single bit $\beta_A$ or $\beta_B$, which indicates that a bock is mined on the given chain when the bit is flipped from one state to the next. 

\medskip

\noindent \textbf{Actions.} Each action $a$ corresponds to the amount of hash rate allocated to chain $A$ for the next second. The remaining hash rate $b$ is allocated to chain $B$ so that $a + b = 6$, i.e. the total hash rate is equal to 6 hashes per second.

\medskip

\noindent \textbf{Transitions.} A transition corresponds to an update in difficulty and block mining status, it occurs any time one of those values change on either blockchain. The only valid change in difficulty is to move from $D_X$ to $D_X' = 2 x$, for $X \in \{A, B\}$, which indicates that when a block is mined the difficulty updates to twice the hash rate applied to the chain (to enforce 2 second block times). For $X \in \{A, B\}$, let $P[D_X, x]$ be the probability that a block is mined on chain $X$ after 1 second, for the given difficulty and hash rate. Similarly, define $C[D_X, x]$ to be the quantity of blocks mined on chain $X$ during a 1 second interval, given that at least one block will be mined. We have the following transition probabilities and rewards where \emph{success} on a given chain is defined as the event of mining at least one block.
\begin{itemize}[topsep=0pt,leftmargin=10pt]
\item \textbf{Both success:} $(D_A, D_B, \beta_A, \beta_B) \rightarrow (D_A', D_B', 1-\beta_A, 1-\beta_B)$ \\
\indent Probability: $P[D_A, a] P[D_B, b]$ \\
\indent Reward: $2 C[D_A, a] + C[D_B, b]$
\item \textbf{$A$ success:} $(D_A, D_B, \beta_A, \beta_B) \rightarrow (D_A', D_B', 1-\beta_A, \beta_B)$ \\
\indent Probability: $P[D_A, a] (1-P[D_B, b])$ \\
\indent Reward: $2 C[D_A, a]$
\item \textbf{$B$ success:} $(D_A, D_B, \beta_A, \beta_B) \rightarrow (D_A', D_B', \beta_A, 1-\beta_B)$ \\
\indent Probability: $(1-P[D_A, a]) P[D_B, b]$ \\
\indent Reward: $ C[D_B, b]$
\item \textbf{None success:} $(D_A, D_B, \beta_A, \beta_B) \rightarrow (D_A', D_B', \beta_A, \beta_B)$ \\
\indent Probability: $(1-P[D_A, a]) (1-P[D_B, b])$ \\
\indent Reward: 0
\end{itemize}
For PoW functions where the best known solving algorithm is trial and error, and negligible network latency, the block inter-arrival time is exponentially distributed. Specifically, Ozisik et al.~\cite{Ozisik:july2017} showed that inter-arrival time $T_i$ for block $i$ has distribution $T \sim \texttt{Expon}(T D_i/x)$, where $T$ is the target block time, $D_i$ is the difficulty (expected number of hashes per block), and $x$ is the actual number of hashes performed every $T$ seconds. For  $X \in \{A, B\}$, the success probability on chain $X$ is therefore given by \vspace{-1em}
\[
P[D_X,x] = \int_0^1 f(t) dt,
\]
where $f(t)$ is the PDF of the distribution $\texttt{Expon}(D_X / x)$. That is to say, $P[D_X,x]$ it is given by the cumulative distribution for the exponential from time 0 until 1 second. To derive the expression for block quantity (in excess of 1 block during a 1 second interval) we must first contemplate the time that the first block is mined, $t$, and then the number of additional blocks that will be mined in the $1-t$ remaining seconds. The former is governed by an exponential distribution, while the latter follows a Poisson distribution. We have
\[
C[D_X,x] \approx 1 + \frac{\int_0^1 f(t) \sum_{i=0}^{10} i f'(1-t) dt}{\int_0^1 f(t) dt,},
\]
where again $f(t)$ is the PDF of the distribution $\texttt{Expon}(D_X / x)$ and $f'(t)$ is the distribution $\texttt{Poisson}(t)$. This expression first adds a reward for the first block mined. Next, for each time $t \in [0,1]$, it calculates the expected quantity of blocks from 0 to 10 total (stopping at 10 because larger values are unlikely), weighted by the probability of mining the first block in time $t$. Finally, to condition on the event that a first block is found, the expected block count beyond 1 is normalized by the probability of mining a block in the first second. Notice that we use unit difficulty for all blocks other than the first since the DAA is assumed to be adjust perfectly at that point.

%% file: example.tex
% !TEX root = main.tex

\section{Futures leveraging \texttt{Oracle}}
\label{sec:fut_example}

\begin{myexam}
\label{exam:futures}
Suppose that we wish to introduce fully decentralized \emph{futures contracts} to blockchain $A$ intended to be negotiated between two parties: guarantor $\mathcal{G}$ and beneficiary $\mathcal{B}$. To do so, a smart contract can be developed that leverages $\texttt{Oracle}$. Each futures contract, or \emph{future} transfers from guarantor to beneficiary a quantity of coins $A$ equivalent to the value of a quantity of coin $B$ at a future date. 
Specifically, a future issued at the time when chains $A$ and $B$ are at block heights $b_A$ and $b_B$, allows the beneficiary to trade the contract to the guarantor for a quantity of coins $A$ equivalent to 1 coin $B$ on the \emph{expiry date}. We define expiry as the latter of block heights $b'_A$ and $b'_B$, anticipated to be some time in the future (for example 90 days). Contract $\texttt{Future}$ implements four methods: $\texttt{Deposit}(a)$, $\texttt{Recover}(a)$, $\texttt{Issue}(b_A, b_B, b'_A, b'_B, a)$, and $\texttt{Redeem}(b_A', b_B')$. $\texttt{Deposit}$ is signed by $\mathcal{G}$; it deposits quantity $a$ of coin $A$ into $\texttt{Future}$. This will be used to pay $\mathcal{B}$ at expiry. Prior to calling $\texttt{Issue}$, the funds can be redeemed by $\mathcal{G}$ if he signs $\texttt{Recover}$. The call to $\texttt{Issue}$ must be signed by both $\mathcal{G}$ and $\mathcal{B}$; signifying that they agree to the initial and expiry block times and fee of $a$ coins, which is paid by $\mathcal{B}$ and immediately transferred to an account owned by $\mathcal{G}$. Once headers $h'_A$ and $h'_B$ at height $b'_A$ and $b'_B$ have been generated, $\mathcal{B}$ first calls $\texttt{Update}(h'_B)$ on $\texttt{Oracle}$ and then signs $\texttt{Redeem}$. In response to this method, contract $\texttt{Future}$ deposits into an account controlled by $\mathcal{B}$ a quantity of $A$ coins that are equivalent to the value of 1 coin $B$ as determined by calling $\texttt{Query}(b_A', b_B')$ on contract $\texttt{Oracle}$. 
\end{myexam}

%% file: algorithms.tex
% !TEX root = main.tex

\vspace{-1em}
\section{Price Oracle Algorithms}\vspace{-0em}
\label{sec:algorithms}

\begin{algorithm}[htp]
\SetAlgoNoLine
 \caption{$\texttt{Oracle}.\texttt{Update}(h_B)$}
 \label{alg:oracle_update}
 \If{$h_B[P] \neq \mathcal{H}(h_B')$}{
 	\Return\;
 }
 \If{$\mathcal{H}(h_B) > h_B'[g]$}{ 
 	\Return\;
 }
 $\texttt{Headers}_B.\texttt{append}(h_B)$\;
\end{algorithm}

\begin{algorithm}[htp]
\SetAlgoNoLine
 \caption{$\texttt{Oracle}.\texttt{Query}(b_A, b_B, \sigma_\Delta)$}
 \label{alg:oracle_query}
  \If{$\texttt{length}(\texttt{Headers}_A) < b_A$}{
   \textbf{throw error}\;
 }
 \If{$\texttt{length}(\texttt{Headers}_B) < b_B$}{
   \textbf{throw error}\;
 }
 \Return $\sigma_\Delta \frac{k_A}{k_B} \frac{\texttt{Headers}_B[b_B][D]}{\texttt{Headers}_A[b_A][D]}$\;
\end{algorithm}

\begin{algorithm}[htp]
\SetAlgoNoLine
 \caption{$\texttt{Spot}.\texttt{Solve}(a, \texttt{n})$}
 \label{alg:spot_solve}
  $P = \mathcal{H}(\texttt{Headers}_A[-1])$\;
  \If{$\mathcal{H}(P, \texttt{n}, a) < g$}{
    $b_A = \texttt{length}(\texttt{Headers}_A)$\;
    $(r, g', g_\Delta') = (1, g, g_\Delta)$\;
    $g_\Delta = g / j$\;
    $g = \alpha g$\;
    $a \mathrel{+}= \mathcal{R}$\;
  }
\end{algorithm}

\begin{algorithm}[h]
\SetAlgoNoLine
 \caption{$\texttt{Spot}.\texttt{Query}(f)$}
 \label{alg:spot_query}
  $\mathcal{F} \mathrel{+}= f$\;
  \Return $\frac{k_A g_A}{\mathcal{R} (\alpha g' + r g_\Delta')}$\;
\end{algorithm}

\begin{algorithm}[h]
\SetAlgoNoLine
 \caption{$\texttt{Spot}.\texttt{Update}()$}
 \label{alg:spot_update}
  \If{$\texttt{length}(\texttt{Headers}_A) > b_A + N$}{
    $b_A = \texttt{length}(\texttt{Headers}_A)$\;
    $r \mathrel{+}= 1$\;
    $g \mathrel{+}= g_\Delta$\;
  }
\end{algorithm}

%% file: proofs.tex
% !TEX root = main.tex

\vspace{-1em}
\section{Proofs}\vspace{-1em}
\label{sec:proofs}

\begin{flushleft}
%%%%%%%%%%% Nash Equilibrium %%%%%%%%%%
\fbox{\parbox{\columnwidth}{{\noindent \textbf{THEOREM \ref{thm:nash_equil}:}\em}
\emph{
The following allocation is a symmetric equilibrium for the Security Allocation Game: 
\[
[\boldsymbol{w}^*_i, \boldsymbol{w}^*_{\minus i}] = \left[ \frac{1}{N} (c, 1-c), \frac{n}{N} (c, 1-c) \right], 
\]
where $n = N-1$ and $c = \frac{T_B R}{T_B R - T_A R + T_A}$. When $T_A = T_B$ the equilibrium simplifies to $c = R$.
}}}

\smallskip
\begin{myproof}
Allocation $[\boldsymbol{w}^*_i, \boldsymbol{w}^*_{\minus i}]$ constitutes a Nash equilibrium if every miner's best response at that point is to maintain the same allocation. Because miner resources are assumed to be homogenous, it will suffice to show that $w_{iA} = \frac{c}{N}$ is the best response when $w_{\minus iA}^* = \frac{cn}{N}$.

When the allocation for miners $m_{\minus i}$ is $w_{\minus iA}$, the best response for miner $m_i$ is given by $\boldsymbol{\pi}^T \boldsymbol{w}_i$, which we denote in this proof simply by $y_i$. 
Thus, our task is to show that $w_{iA}^* = \frac{c}{N}$ is the global optimum of $y_i$ when $w_{\minus iA} = \frac{cn}{N}$. To that end, we proceed by identifying and testing the critical points of function $y_i$, beginning with its local optima. 

Solving $\frac{\partial y_i}{\partial w_{iA}} = 0$ gives all local optima. We have
\[
\frac{w_{\minus iA}}{(w_{iA} + w_{\minus iA})^2}  \frac{V_A}{T_A} + \frac{w_{\minus iA} - \frac{n}{N}}{(1 - w_{\minus iA} - w_{iA})^2} \frac{V_B}{T_B} = 0,
\]
which implies
\[
\begin{array}{rcl}
&& \frac{w_{\minus iA}}{(w_{iA} + w_{\minus iA})^2}  \frac{V_A}{T_A} = \frac{\frac{n}{N} - w_{\minus iA}}{(1 - w_{\minus iA} - w_{iA})^2} \frac{V_B}{T_B} \\
&\Rightarrow& \frac{\sqrt{w_{\minus iA}}}{w_{iA} + w_{\minus iA}}  \sqrt{\frac{V_A}{T_A}} = \pm \frac{\sqrt{\frac{n}{N} - w_{\minus iA}}}{1 - w_{\minus iA} - w_{iA}} \sqrt{\frac{V_B}{T_B}}.
\end{array}
\]
The quantity on the left is always positive and because $w_{\minus iA} < \frac{n}{N}$ and $w_{iA} + w_{\minus iA} < 1$, the absolute value of the quantity on the right is also positive. Therefore, only the positive branch of the square root leads to a valid solution. It follows that 
\begin{equation}
\label{eq:best_response}
w_{iA} = \frac{(1 - w_{\minus iA}) \sqrt{w_{\minus iA} V_A T_B} - w_{\minus iA} \sqrt{(\frac{n}{N}-w_{\minus iA})V_B T_A}}{\sqrt{w_{\minus iA} V_A T_B} + \sqrt{(\frac{n}{N} - w_{\minus iA}) V_B T_A}}
\end{equation}
is the only local optimum. Thus, we proceed by performing the substitution $w_{\minus iA}^* = \frac{c n}{N}$ in Eq.~\ref{eq:best_response} and showing that its value is equal to $\frac{c}{N}$. Using $\frac{1-c}{c} = \frac{V_B}{V_A} \frac{T_A}{T_B}$ and $c = \frac{T_B V_A}{T_B V_A + T_A V_B}$, we have  (algebra not shown)
\[
\begin{array}{rcl}
w_{iA} &=& \frac{(1 - \frac{c n}{N}) \sqrt{\frac{c n}{N} V_A T_B} - \frac{c n}{N} \sqrt{(\frac{n}{N}-\frac{c n}{N})V_B T_A}}{\sqrt{\frac{c n}{N} V_A T_B} + \sqrt{(\frac{n}{N} - \frac{c n}{N}) V_B T_A}} \\ [+8pt]
&=& \frac{c}{N}.
\end{array}
\]
The payoff to miner $i$ for allocation $c \left( \frac{1}{N}, \frac{n}{N} \right)$ is
\[
\begin{array}{l}
y_{i c} = 
\left( \frac{\frac{c}{N}}{\frac{c}{N} + \frac{cn}{N}} \frac{V_A}{T_A} + \frac{\frac{(1-c)}{N}}{\frac{(1-c)}{N} + \frac{(1-c)n}{N}} \frac{V_B}{T_B} \right)  \\
=\frac{1}{N} \left( \frac{V_A}{T_A} + \frac{V_B}{T_B} \right).
\end{array}
\]

Next, we turn our attention to proving that $[\boldsymbol{w}^*_i, \boldsymbol{w}^*_{\minus i}]$ is actually a global optimum by showing its payoff, $y_{i c}$, exceeds that of other critical points of the payoff function. Endpoints 
$[(0, \frac{c}{N}), (\frac{cn}{N}, 0)]$ and $[(\frac{c}{N}, 0), (\frac{cn}{N}, 0)]$ 
constitute the remaining critical points. Their payoffs are, respectively,
\[
y_{i0} = \frac{1}{N - cn} \frac{V_B}{T_B} \mbox{~~and~~}
y_{i\frac{1}{N}} = \frac{1}{1+cn} \frac{V_A}{T_A}.
\]
It can be shown that $y_{i c} \geq y_{i0}$ and $y_{i c} \geq y_{i\frac{1}{N}}$ for all choices of $\frac{V_A}{T_A}$ and $\frac{V_B}{T_B}$. Therefore, allocation $w_{iA} = \frac{c}{N}$ maximizes payoff when $w_{\minus iA} = \frac{cn}{N}$, so allocation $[\boldsymbol{w}^*_i, \boldsymbol{w}^*_{\minus i}]$ is a Nash equilibrium.
\end{myproof}

%%%%%%%%%%% No Arbitrage %%%%%%%%%%%%%

\fbox{\parbox{\columnwidth}{{\noindent \textbf{PROPOSITION \ref{thm:portfolio_price}:}\em}
\emph{
For any fixed allocation $\boldsymbol{w}$, after the SAAs on chains $A$ and $B$ come to rest, the  portfolio pricing vector will be $\boldsymbol{p} = e \boldsymbol{w}$.
}}}

\medskip

\begin{myproof} Consider a blockchain $X$ with aggregate claim $c_X(\tau)$ and prevailing price $p_X(\tau) = S_X(\tau)$. Together these two quantities entirely determine the actual security investment applied to the chain: 
\begin{equation}
s_X(\tau) = c_X(\tau) p_X(\tau) = c_X(\tau) S_X(\tau). 
\end{equation}
Thus in order for the SAA to be at rest, it must be the case that $c_X(\tau) = 1$. The same reasoning can be applied to any blockchain, so that if both SAAs are at rest at time $\tau$, then $\boldsymbol{c}(\tau)$ is a vector of all ones. Finally, from Eq.~\ref{eq:allocation} we have that 
\[
\boldsymbol{p}(\tau) = e \boldsymbol{w}(\tau) \oslash \boldsymbol{c}(\tau) =  e \boldsymbol{w}(\tau).
\]
\end{myproof}

\fbox{\parbox{\columnwidth}{{\noindent \textbf{THEOREM \ref{thm:equil}:}\em}
\emph{
Assume any choice of SAA for chains $A$ and $B$ (not necessarily the same). When the relative reward $R$ is stable, there exists no arbitrage at the following allocation 
\[
\boldsymbol{w}_{\texttt{eq}} = \left( \frac{T_B R}{T_B R - T_A R + T_A}, \frac{T_A (1 - R)}{T_B R - T_A R + T_A} \right),
\]
which simplifies to 
\[
\boldsymbol{w}_{\texttt{eq}} = (R, 1 - R),
\]
if $T_A = T_B$.
}}}

\medskip
\begin{myproof}
Suppose that the current allocation is $\boldsymbol{w}_{\texttt{eq}}$ and both SAAs are at rest. From Proposition~\ref{thm:portfolio_price}, we know that the price of claims is given by
\[
\boldsymbol{p}_{\texttt{eq}} = \frac{e}{T_B R - T_A R + T_A} (T_B R, T_A(1-R)).
\]
Now consider the payoff and price associated with the change in claim that manifests the following change in allocation:
\[
\Delta \boldsymbol{w} = \frac{1}{T_B R - T_A R + T_A} (\delta_1, -\delta_2),
\]
for arbitrary $\delta_1, \delta_2 > 0$. That is to say, the allocation to chain $A$ is boosted proportional to $\delta_1$ while the allocation to chain $B$ is sold short proportional to $\delta_2$. According to Eq.~\ref{eq:allocation}, in the moments before either SAA responds to this allocation change, the claim associated with $\Delta \boldsymbol{w}$ becomes $\Delta \boldsymbol{c} = e \Delta \boldsymbol{w} \oslash \boldsymbol{p}_{\texttt{eq}}$, or
\[
\Delta \boldsymbol{c} = \left(\frac{\delta_1}{T_B R}, \frac{-\delta_2}{T_A(1-R)} \right).
\]
Therefore, the payoff associated with this change is
\[
\Delta \boldsymbol{c}^T \boldsymbol{\pi} = \frac{\delta_1 V_A}{T_A T_B R} -\frac{\delta_2 V_B}{T_A T_B(1-R)} = (\delta_1 - \delta_2) \frac{(V_A + V_B)}{T_A T_B}.
\]
And the corresponding price is
\[
\Delta \boldsymbol{c}^T \boldsymbol{p} = \frac{e(\delta_1 - \delta_2)}{T_B R - T_A R + T_A} =  (\delta_1 - \delta_2) \frac{e (V_A + V_B)}{T_B V_A + T_A V_B}.
\]

To prove the theorem, it will suffice to show that \1 when $\Delta \boldsymbol{c}^T \boldsymbol{\pi}> 0$, $\Delta \boldsymbol{c}^T \boldsymbol{p} \geq 0$ and \2 when $\Delta \boldsymbol{c}^T \boldsymbol{p} < 0$, $\Delta \boldsymbol{c}^T \boldsymbol{\pi} \leq 0$. To that end, note that in order for $\Delta \boldsymbol{c}^T \boldsymbol{\pi} > 0$, it must be the case that $\delta_1 > \delta_2$. Therefore, $\Delta \boldsymbol{c}^T \boldsymbol{p} > 0$. Conversely, if $\Delta \boldsymbol{c}^T \boldsymbol{p} < 0$, then $\delta_2 > \delta_1$, which implies that $\Delta \boldsymbol{c}^T \boldsymbol{\pi} < 0$.
\end{myproof}

\fbox{\parbox{\columnwidth}{{\noindent \textbf{LEMMA \ref{eq:sym_rebalance}:}\em}
\emph{
For initial allocation $\boldsymbol{w}$ and price $\boldsymbol{p}$, the claims associated with a symmetric rebalancing  $\Delta \boldsymbol{w}$ are given by $\Delta \boldsymbol{c} = \Delta \boldsymbol{w} \oslash \boldsymbol{w}$ and it is always the case that $\Delta \boldsymbol{c}^T  \boldsymbol{p} = 0$.
}}}

\medskip
\begin{myproof}
According to Eq.~\ref{eq:allocation}, prior to either SAA responding to the allocation rebalancing, the claim associated with $\Delta \boldsymbol{w}$ is given by \vspace{-1em}
\begin{equation}
\label{eq:delta_c}
\Delta \boldsymbol{c} = e \Delta \boldsymbol{w} \oslash \boldsymbol{p}. 
\end{equation}
Meanwhile, Proposition~\ref{thm:portfolio_price} establishes that $\boldsymbol{p} = e  \boldsymbol{w}$. Thus, it follows that $\Delta \boldsymbol{c} = \Delta \boldsymbol{w} \oslash \boldsymbol{w}$. Returning to Eq.~\ref{eq:delta_c}, it is also apparent that $\Delta \boldsymbol{c}^T  \boldsymbol{p} = e (\Delta w_A + \Delta w_B)$, which is always zero provided that $\Delta \boldsymbol{w}$ is symmetric.
\end{myproof}

%\newpage
\fbox{\parbox{\columnwidth}{{\noindent \textbf{THEOREM \ref{thm:na_unique}:}\em}
\emph{
For any allocation $\boldsymbol{w} \neq \boldsymbol{w}_{\texttt{eq}}$, there exists a symmetric allocation rebalancing $\Delta \boldsymbol{w}$, such that $|(\boldsymbol{w} + \Delta \boldsymbol{w}) - \boldsymbol{w}_{\texttt{eq}}| \leq |\boldsymbol{w} - \boldsymbol{w}_{\texttt{eq}}|$, which has price zero and strictly positive payoff.
}}}

\medskip
\begin{myproof}
Without loss of generality we may assume that $w_A < w_{\texttt{eq}A}$, which implies that $\boldsymbol{w} = \boldsymbol{w}_{\texttt{eq}} - (\delta_1, -\delta_2)$ for $\delta_1$ and $\delta_2$ such that $0 < \delta_1, \delta_2 < 1$. Let $\Delta \boldsymbol{w} = (\epsilon, -\epsilon)$ for some $\epsilon < \min\{\delta_1, \delta_2\}$. Note that, by construction, $|(\boldsymbol{w} + \Delta \boldsymbol{w}) - \boldsymbol{w}_{\texttt{eq}}| \leq |\boldsymbol{w} - \boldsymbol{w}_{\texttt{eq}}|$.
According to Lemma~\ref{eq:sym_rebalance}, we have
\[
\Delta \boldsymbol{c} = \Delta \boldsymbol{w} \oslash (\boldsymbol{w}_{\texttt{eq}} - (\delta_1, \delta_2)).
\]
Substituting values for $\Delta \boldsymbol{w}$ and $\boldsymbol{w}_{\texttt{eq}}$ yields
\[
\Delta \boldsymbol{c} = \left( \frac{\epsilon (T_B V_A + T_A V_B)}{(1-\delta_1) T_B V_A - \delta_1 T_A V_B}, \frac{-\epsilon (T_B V_A + T_A V_B)}{(1+\delta_2) T_A V_B + \delta_2 T_B V_A} \right).
\]
It follows that payoff is given by
\[
\begin{array}{l}
\Delta \boldsymbol{c}^T \boldsymbol{\pi} = 
 \alpha \left( \frac{V_A}{(1-\delta_1) T_A T_B V_A - \delta_1 T_A^2 V_B} - \frac{V_B}{(1+\delta_2) T_A T_B V_B + \delta_2 T_B^2 V_A} \right),
\end{array}
\]
where $\alpha = \epsilon (T_B V_A + T_A V_B)$. Notice that $(1-\delta_1)T_B V_A - \delta_1 T_A V_B = w_{\texttt{eq}A} - \delta_1 > 0$. Therefore, both terms in the difference above are positive. It follows that payoff $\Delta \boldsymbol{c}^T \boldsymbol{p}$ will be greater than zero provided that
\[
-\delta_1 T_A^2 V_B^2 - \delta_2 T_B^2 V_A^2 < (\delta_1 + \delta_2) T_A T_B V_A V_B,
\]
which is true for all valid $\delta_1$ and $\delta_2$.

Next, consider the price of rebalancing: $\Delta \boldsymbol{c}^T \boldsymbol{p}$. Since $\Delta \boldsymbol{c} = e \Delta \boldsymbol{w} \oslash \boldsymbol{p}$, it follows that 
\[
\Delta \boldsymbol{c}^T \boldsymbol{p} = e (\Delta w_A + \Delta w_B) = e(\epsilon - \epsilon) = 0,
\]
which implies that the price associated with rebalancing is zero.
\end{myproof}

\fbox{\parbox{\columnwidth}{{\noindent \textbf{COROLLARY \ref{cor:na_unique}:}\em}
\emph{
For allocation $\boldsymbol{w} \neq \boldsymbol{w}_{\texttt{eq}}$, any symmetric rebalancing allocation $\Delta \boldsymbol{w}$ such that $|(\boldsymbol{w} + \Delta \boldsymbol{w}) - \boldsymbol{w}_{\texttt{eq}}| > |\boldsymbol{w} - \boldsymbol{w}_{\texttt{eq}}|$ has price zero will result in strictly negative payoff.
}}}

\medskip
\begin{myproof}
Again, without loss of generality, we may assume that $w_A < w_{\texttt{eq}A}$ and that $\boldsymbol{w} = \boldsymbol{w}_{\texttt{eq}} - (\delta_1, -\delta_2)$. To ensure that $|(\boldsymbol{w} + \Delta \boldsymbol{w}) - \boldsymbol{w}_{\texttt{eq}}| > |\boldsymbol{w} - \boldsymbol{w}_{\texttt{eq}}|$, it must be the case that $\Delta \boldsymbol{w} = (-\epsilon, \epsilon)$ for some $\epsilon > 0$. 
Since $\Delta \boldsymbol{w}$ is symmetric, Lemma~\ref{eq:sym_rebalance} ensures that it achieves a portfolio price of zero. Following closely to the derivation in Theorem~\ref{thm:na_unique}, the payoff is given by
\[
\begin{array}{l}\vspace{-.2em}
\Delta \boldsymbol{c}^T \boldsymbol{\pi} =  
\left( \frac{-\epsilon V_A (T_B V_A + T_A V_B)}{(1-\delta_1) T_A T_B V_A - \delta_1 T_A^2 V_B} + \frac{\epsilon V_B (T_B V_A + T_A V_B)}{(1+\delta_2) T_A T_B V_B + \delta_2 T_B^2 V_A} \right),
\end{array}
\]
which can never be positive.\vspace{-.5em}
\end{myproof}

\end{flushleft}